\documentclass[journal]{IEEEtran}                         

\IEEEoverridecommandlockouts                              

\usepackage{amsfonts}
\usepackage{mathrsfs}
\usepackage{amssymb}
\usepackage{amsmath}
\usepackage{mathrsfs}
\usepackage{array}
\usepackage{extarrows}
\usepackage{color, xcolor}
\usepackage{cite}
\usepackage{graphicx}
\usepackage{float}
\usepackage{mathtools}
\usepackage[font=footnotesize]{caption}
\usepackage{stfloats}
\usepackage{epstopdf}
\usepackage{arydshln}
\usepackage{ifthen}
\usepackage{url}
\usepackage{pict2e}

\def\rep#1{(\ref{#1})}

\newcommand{\R}{\mathbb{R}}

\def\send#1#2{\stackrel{#1}{\hbox to #2{\rightarrowfill}}}
\def\-{\!\!\!\!\!-}

 \def\qed{ \rule{.1in}{.1in}}
\def\eq#1{\begin{equation}#1\end{equation}}

\newcommand{\rank}{{\rm rank\;}}

\def\scr#1{{\cal #1}}

\newcommand{\dfb}{\stackrel{\Delta}{=}}

\newtheorem{theorem}{Theorem}
\newtheorem{lemma}{Lemma}
\newtheorem{remark}{Remark}
\newtheorem{proposition}{Proposition}
\newtheorem{corollary}{Corollary}

\def\qed{ \rule{.1in}{.1in}}

\def\R{{\rm I\!R}} 

\newcounter{seqn}[equation]
\def\theseqn{\arabic{equation}\alph{seqn}}

\def\endseqn{\eqno \@seqnnum
$$\ignorespaces}
\def\@seqnnum{(\theseqn)}

\newskip\mcentering \mcentering=0pt plus 1000pt minus 1000pt

\def\meqalignno#1{
\halign to\displaywidth{
    \hbox to 0pt{\kern\displaywidth\llap{$##$}\hss}\tabskip=\mcentering
    &\hfil$\displaystyle{##}$\tabskip=\mcentering
   &&$\displaystyle{{}##}$\hfil\tabskip=\mcentering
    \crcr
    #1\crcr}}

\def\rep#1{(\ref{#1})}

\def\eq#1{\begin{equation}#1\end{equation}}

\def\dspace{\multiply\normalbaselineskip 150
		  \divide\normalbaselineskip 100 \normalbaselines
		  \csname @@normalbaselineskip\endcsname\normalbaselineskip}
\def\sspace{\multiply\normalbaselineskip 200
		 \divide\normalbaselineskip 300 \normalbaselines
		 \csname @@normalbaselineskip\endcsname\normalbaselineskip}
\def\sdspace{\multiply\normalbaselineskip 160
		 \divide\normalbaselineskip 150 \normalbaselines
		 \csname @@normalbaselineskip\endcsname\normalbaselineskip}

\def\@{\tilde}

\def\3dot#1{\buildrel\textstyle...\over#1}

\title{\LARGE \bf
Distributed Feedback Control of Multi-Channel Linear Systems
}

\author{F. Liu, L. Wang, D. Fullmer, and A. S. Morse
\thanks{This work was supported by National Science Foundation grant n. 1607101.00, US Air Force grant n.
FA9550-16-1-0290, and Army Research Office grant n. W911NF-17-0499.}
\thanks{
F. Liu is with the School of Aerospace Engineering, Georgia Institute of Technology, Atlanta, GA 30332 USA (e-mail: fengjiao@gatech.edu).
L. Wang is with the Department of Electrical Engineering and Computer Science, University of California, Irvine, Irvine, CA 92618 USA (e-mail: lili.wang.zj@gmail.com). 
D. Fullmer may be reached using e-mail: danielrf12@gmail.com.
A. S. Morse is with the Department of Electrical Engineering, Yale University, New Haven, CT 06511 USA (e-mail: as.morse@yale.edu). 
Corresponding author: L. Wang.}
\thanks{Portions of this paper were presented in abbreviated forms at the 2020 American Control Conference
 \cite{wang2020distributed}
and at the 2020 IEEE Conference on Decision and Control \cite{liu2020what}.
 An overview of the  material in \S \ref{sec:dis-obs-contr} and \S \ref{sec:set-point} without proofs
can be found in \cite{wang2020distributed} whereas a brief version of the material in \S \ref{liu} and
 \S \ref{sec:delays}, also without proofs,
can be found in \cite{liu2020what}.
 The reader interested in highlights of this paper without the details is directed to these two conference
 reprints.}
        }

\begin{document}
\maketitle

\thispagestyle{plain} 
\pagestyle{plain}     

\begin{abstract}
In this paper it is  established  that any jointly controllable, jointly observable,
 multi-channel, discrete or continuous time linear system with a strongly connected neighbor \{communication\}
 graph can be exponentially stabilized with any pre-specified   convergence rate
 using a time-invariant distributed  linear control. As an illustration of how this
  finding
 can be used to deal with certain distributed tracking problems,
 a  solution is given
  to a distributed
  set-point control problem for a continuous-time, multi-channel linear system
   in which each
   and every agent with access to the system is able to independently adjust its
  scalar-valued,   controlled output to any desired set-point value.
  To better understand the constraints on controller design, the
  distributed control problem  is recast  as a classical decentralized control problem.
Armed with the tools of decentralized control, including the notion of a ``fixed spectrum'',
it is   possible to show quite surprisingly  that if the only information each agent is allowed
  to share  with its neighbors is its measured output, then
  distributed stabilization in some cases is   impossible.
 Using well-known decentralized control concepts, lower bounds are derived   on the  dimensions of the shared sub-states
 of local controllers which, if satisfied,
  guarantee that   there will be no fixed closed-loop system eigenvalues to contend with.
  The decentralized control perspective also enables one to   assert definitively  that without imposing a partitioning constraint,
  the closed-loop spectrum of any
  jointly controllable, jointly observable multi-channel linear system with a strongly connected neighbor graph,
  can be freely assigned with
  distributed  feedback control.
The paper then turns to the
 important and  often overlooked design problem of dealing with the effects of transmission delays across a network.
      It is explained why in the face of finite delays, exponential stabilization at any
      prescribed convergence rate can still be achieved with distributed control, at least for
       discrete-time multi-channel linear systems.

\end{abstract}

\begin{IEEEkeywords}
Multi-channel linear systems, distributed stabilization, fixed eigenvalues, transmission delays.
\end{IEEEkeywords}

\vspace{-0.3cm}

\section{Introduction}\label{intro}

 Distributed control and estimation have been
 under active study for more than twenty years.
The central aim of this paper is to contribute to this technology by solving
 what is arguably
 the most fundamental problem in distributed control, namely to constructively prove that any
 ``multi-channel'' linear system with a strongly connected ``neighbor'' \{communication\}
 graph can be exponentially stabilized with an arbitrarily fast convergence rate
 using a time-invariant distributed  linear control.
By an $n$-dimensional,  {\em multi-channel},  continuous-time, linear system with $m$ channels
is meant a linear system of the form
\begin{equation} \label{mcs}
\dot{x} = Ax + \sum_{i=1}^m B_i u_i, \hspace{5mm} y_i = C_i x, \hspace{5mm} i \in \mathbf{m}
\end{equation}
where $n$ and $m$ are positive integers, $\mathbf{m} \triangleq \{1, 2, \dots, m\}$, $x \in {\rm I\!R}^n$,
 and for each $i \in \mathbf{m}$, $u_i \in {\rm I\!R}^{p_i}$ is
  the control input of channel $i$ and $y_i \in {\rm I\!R}^{q_i}$ is
   the measured output of channel $i$. Here $A$, $B_i$, and $C_i$ are real-valued,
    constant matrices of appropriate sizes. The discrete-time counterpart of (\ref{mcs})
     is an $m$-channel linear system of the form
\eq{x(t+1) = Ax(t) + \sum_{i=1}^m B_i u_i(t), \;\;
y_i(t) = C_i x(t), \; i \in \mathbf{m}\label{mcsd}}
where $t \in \{0, 1, 2, \dots\}$.
It is presumed that either system (\ref{mcs}) or
 (\ref{mcsd}) is to be controlled by $m$ agents, labeled $1$ through $m$,
  with the understanding that each agent $i \in \mathbf{m}$ can measure output
  signal $y_i$ and has access to control input $u_i$. It
   is further assumed that each agent $i$ can  receive the measured output of each of its ``neighbors'' as well  as some
suitably defined sub-state  of each of its neighbors' controllers.
The specification of who agent $i$'s neighbors are is part of the problem formulation.
The set of labels of agent $i$'s neighbors, including itself, is denoted by $\mathcal{N}_i$.
      The \emph{neighbor graph} associated with either \rep{mcs} or \rep{mcsd}, written $\mathbb{N}$, is a directed graph on $m$
      vertices, with an arc from vertex $j$ to vertex $i$ just in case agent $j$ is a neighbor of
       agent $i$. It is assumed that each agent's neighbors do not change with time.
       Thus $\mathbb{N}$ is a stationary graph.

\vspace{-0.4cm}
\subsection{The Problem}\label{bdcp}
\vspace{-0.1cm}

It is assumed  throughout this paper that the multi-channel system under consideration is both
 jointly controllable and jointly observable; that is for  the
system defined by
\rep{mcs} or \rep{mcsd}  the matrix pairs,
\begin{equation*}
\left(
A, \enspace
\left[
B_1 \enspace B_2 \enspace \cdots \enspace B_m
\right]
\right)
\hspace{2mm} \text{and} \hspace{2mm}
\left(
\begin{bmatrix}
C_1'&
C_2'&
\ldots &
C_m'
\end{bmatrix}'
, \enspace A
\right)
\end{equation*}
are controllable and observable, respectively, where $'$ denotes transposition.  
    For simplicity, it is also assumed that for $i\in\mathbf{m}$, both $B_i\neq 0$
  and
 $C_i\neq 0$. Subject to these assumptions,
   the main problem  to which this paper is addressed is as follows.

\noindent{\em Basic Distributed Control Problem:} For the $m$-channel system defined by
(\ref{mcs}) or (\ref{mcsd}),  develop  a systematic procedure for deciding which signals each agent is to receive from its neighbors and for
constructing $m$ linear time-invariant feedback controllers using these signals, one for each
channel, so that the state of the resulting closed-loop system converges
to zero exponentially fast at a pre-assigned rate.

\vspace{-0.4cm}
\subsection{Summary}
\vspace{-0.1cm}

In this paper it is shown  that any jointly controllable, jointly observable,
 multi-channel, discrete or continuous time linear system with a strongly connected neighbor \{communication\}
 graph can be exponentially stabilized with an arbitrarily fast convergence rate
 using a time-invariant distributed  linear control. This is proved constructively
 in \S \ref{sec:dis-obs-contr} for  continuous-time systems
  using a distributed
  observer-based certainty
 equivalence control. Here's how this is done. The notion of   certainty equivalence control is briefly over-viewed in \S \ref{CEC} along with
   explanations of what the ``cancellation'' and ``substitution'' rules are and why they are important.
A brief review is then  given in \S \ref{pop}
 of the ``open-loop'' structure of the distributed observer \cite{wang2017distributed} which will be used.
Finally it is   explained in \S \ref{inputs} how to
modify this observer  so that it can be used in a feedback configuration. What results is a distributed
 observer-based architecture which solves the distributed control problem of
  interest for systems with strongly connected neighbor graphs.
The same construction applies  to  discrete-time systems.
  As an illustration of how these ideas can be
 used to deal with certain distributed tracking problems,
 a  solution is then  given in \S\ref{sec:set-point}
  to a distributed
  set-point control problem for a continuous-time, multi-channel linear system
   in which each
   and every agent with access to the system is able to independently adjust its
  scalar-valued,   controlled output to any desired set-point value.

With the aim of understanding more clearly some of the
 implications on controller design  of the distributional constraint,
  the  distributed control problem of interest is recast in an algorithmically independent way,  as a
  classical decentralized control problem \{\S\ref{liu}\}. This is done
   by first reviewing in \S\ref{dcontrol}, needed results from
classical decentralized control theory \cite{wangdavison, corfmat1976decentralized} including the notion of a ``fixed spectrum''.
Armed with these concepts   it is   possible to show  that if the only information each agent is allowed
  to share  with its neighbors is its measured output, then
  distributed stabilization in some cases is   impossible \{\S\ref{nope}\}.
This surprising observation is universal. It holds for all possible linear time-invariant  distributed controls
 one might consider for either
   discrete or continuous time systems.
The concept of an ``extended system'' is then introduced
 in \S\ref{esyst} and  used in
 \S \ref{restate} to restate the basic distributed
control problem for the original system as a decentralized control problem for the extended system. Doing this enables one
to address the distributed control problem
without explicitly  appealing to any one particular type of distributed dynamic compensator such as those discussed in this paper
and  in \cite{kexin,plug,Luis,XZ}.
Viewing the distributed control problem as a decentralized control problem has several immediate payoffs. For example,
it enables one to easily derive lower bounds  on the  dimensions of the shared sub-states of local controllers which, if satisfied,
  guarantee that there will be no fixed closed-loop system eigenvalues to contend with \{cf. Theorem \ref{thm:strg-graph}\}.
 It also enables one to derive a corollary to this theorem
  which  asserts definitively  that without imposing a partitioning constraint, the closed-loop spectrum of any
  jointly controllable, jointly observable multi-channel linear system with a strongly connected neighbor graph, can be freely assigned with
  distributed  feedback control \{\S \ref{freedom}\}.

Finally, the paper  turns to the
 important and  often overlooked design problem of dealing with the effects of transmission delays across the network.
      It is explained in \S \ref{sec:delays}  why in the face of finite delays, exponential stabilization at any
      prescribed convergence rate can still be achieved with distributed control. This  unexpected  observation applies to
       discrete-time multi-channel linear systems.

\vspace{-0.4cm} 
\subsection{Background}
\vspace{-0.1cm} 

There are several   papers directly concerned   with the basic distributed
control problem  formulated in \S \ref{bdcp}.
Among these  is \cite{kexin} which explains how to construct a time-invariant distributed controller capable of
exponentially  stabilizing  a  continuous-time multi-channel  system under the same assumptions made in this paper.
This appears to be the first result of its kind and is noteworthy. A similar time-invariant controller has recently
 been described in the first part of \cite{plug}.
There are several significant differences between what is done in  \cite{kexin,plug}  and what is done in this paper.
For example, while the controllers developed in
\S \ref{sec:dis-obs-contr} of this paper is  ``observer-based,'' those developed in \cite{kexin,plug} are not in that no state-estimation
is carried out and the certainty equivalence idea is not used. In contrast to the controller construction
  described in \S \ref{sec:dis-obs-contr} of this paper,
 the  constructions given in \cite{kexin,plug}  do not provide a means for controlling convergence rate, although it is plausible that they can be so modified;
 also in contrast to this paper,
 the design methodologies proposed in \cite{kexin,plug}  are of the  ``high-gain'' type and  do not have a
  discrete-time counterpart.
High gain controllers can of course become  problematic when issues such as unmodeled dynamics and
measurement noise are taken into account.

One of the biggest stumbling blocks encountered in trying to mimic centralized  observer-based control in a distributed setting
 is that the certainty equivalence
 ``cancellation rule''
\{cf. \S\ref{CEC}\} cannot be satisfied without violating the distributed information pattern assumption.
 The cancellation rule is satisfied in a centralized setting by applying to the state estimator, the same feedback signal that is
 applied to the process. Satisfaction of the cancellation rule ensures that the dynamics of the state estimation error functions autonomously.
 To overcome this limitation, \cite{Luis} proposes a new idea called the ``substitution rule'' \{cf. \S\ref{CEC}\}, which  if satisfied also
    results in an autonomous error system. However unlike the centralized case, in the distributed case  the construction
   of a provably correct distributed observer compatible with this rule  is an  especially challenging problem.  An effort
    to devise such a compatible observer is undertaken in \cite{Luis} by posing the observer's  construction
    as an optimization problem; however, this work is preliminary in that
     no constructive results are presented.  An effort is also made   in \cite{XZ}
     to construct a
      distributed observer compatible with the  substitution rule. This work is also preliminary
      in that no provably correct procedure is described  for constructing  a compatible distributed observer
       except under  restrictive assumptions and conditions.

\vspace{-0.4cm}

\section{Distributed Observer-Based Certainty Equivalence Control} \label{sec:dis-obs-contr}

The aim of this section is to explain how to solve the basic distributed control problem using ``certainty equivalence'' and a
suitably defined distributed observer. We begin with a brief discussion about certainty equivalence itself and about
 the difficulty in applying it in a distributed context.

\vspace{-0.4cm}
\subsection{Certainty Equivalence Control}\label{CEC}

 The idea of certainty
  equivalence   has its roots extending as far  back in 1957 \cite{theil}. The concept has been
 broadened over time
   to encompass not just linear control    but also much of parameter adaptive control
    and even certain classes of nonlinear control. For a process modeled by a
 linear system of the form
   $y=Cx$, $\dot{x} = Ax+Bu$ which is to be regulated by a centralized control,
 the certainty equivalence approach amounts to  first devising   an appropriate
{\em candidate  control} $u=Fx$  assuming the process state $x$ is available for feedback, and then
using instead of $x$ in the definitions of  $u$, a  suitably defined estimate
$\hat{x}$ of $x$ \{i.e. $u = F\hat{x}$\} even though the
 estimate may not be correct. The estimate  is typically generated by an  estimator of the form
 $\dot{\hat{x}} =(A+KC)\hat{x} -Ky +BF\hat{x}$ which may be either an observer or a Kalman filter.
It is important to note that the signal $BF\hat{x}$  is included as an input to
 the estimator to ensure that {\em cancellation rule} is satisfied.
 That is, in forming
 the dynamical system which models the evolution over time of the   state estimation error $e=\hat{x}-x$, the term $BF\hat{x}$ is canceled  and what results
 is the {\em autonomous}
 error system  $\dot{e} = (A+KC)e$. This    has several important consequences. First, the designs of
  the feedback matrix  $F$ and the estimator matrix $K$  can be dealt with  separately; this is a
 manifestation of the so-called ``separation principal'' of stochastic optimal control \cite{separation}. This is
  important because the closed-loop spectrum of the process equals the disjoint union of the spectrum of $A+BF$
   and the spectrum of $A+KC$
   Second, the fact that
  $u = Fx +Fe$ means  if the estimation error
   is ``small,''  the  behavior of the  process is approximately the same as that which
   would have resulted in had the actual candidate control $u=Fx$ been used.

For the distributed control  problem under consideration in this paper,  certainty equivalence has different consequences.
   The certainty equivalence approach for distributed control  starts
off mimicking    the centralized case; i.e.,   a  candidate distributed control   of the form
  $u_i = F_ix$ for each agent $i\in\mathbf{m}$ is crafted  using standard techniques.
Next, instead of using $x$ in the definitions of the  $u_i$, in accordance with certainty equivalence
 agent $i$ uses a suitably defined estimate
$x_i$ of $x$ \{i.e. $u_i = F_ix_i,\;i\in\mathbf{m}$\}.
To carry out this program, it is  of course necessary for there to be available   a provably correct local agent
  estimator
 for generating each  $x_i$.
For continuous-time systems  there are  a handful
 different kinds of ``open-loop'' distributed observers which might be considered for this purpose
 \cite{park2016design,wang2017distributed,kim2016distributed,han2018simple,wang2019distributed-cont,XZ}. No matter which
  one is considered for agent $i$, for the
 cancellation rule to be satisfied, the estimator would have
 to include as an input, the input to the process, just as in the centralized case. However in the distributed case, the process
 input
 is $\sum_{j=1}^{m} B_jF_jx_j$ and adding $\sum_{j=1}^{m} B_jF_jx_j$  as an input to agent $i$'s estimator clearly
 cannot be done without violating distributional assumptions,
  except in the very special case
 when every agent $j$ for which $F_j\neq 0$,  is a neighbor of agent $i$.
 Thus  if all $F_j\neq 0$,  the cancellation rule cannot be satisfied unless the neighbor graph is complete.
  One way to circumvent this problem
  is to add to  agent $i$'s  estimator  the signal
$\sum_{j=1}^{m} B_jF_jx_i$
rather than  $\sum_{j=1}^{m} B_jF_jx_j$. This simple but clever idea, which seems to have been first suggested in \cite{Luis},  is henceforth called the {\em substitution rule}.  Its motivation is pretty clear. 
If each $x_i$ estimates $x$ as intended,
then each $x_i$ also estimates every   $x_j$, $j\in\mathbf{m}$, so there is not much lost in using $x_i$ instead of $x_j$ in agent $i$'s local estimator.
What is nice about  the substitution rule is that it causes the error  system modeling the dynamics of the estimation errors 
$e_i = x_i-x$, $i\in\mathbf{m}$, to be 
autonomous, just as the error model  is in the centralized case. 
Since agent $i$'s input to the process can be written as $u_i = F_ix + F_i e_i$, it means that if the local estimators can be designed to function as intended, $e_i$ will become small and $u_i$  will be approximately the same as agent $i$'s 
candidate control input $F_ix$ which is ultimately what is desired.
But there is a catch. Although adherence to  the  substitution rule causes the error system  to be autonomous,
it also means that the dynamical model for each local estimation  error $e_i$ will have as an input
 the signal $\sum_{j=1}^{m} B_jF_j(e_i - e_j)$ \{cf. \S\ref{inputs}\}.
No matter which distributed observer cited above  is used,  these signals will always be present in the error equations
 and their presence  makes the design of the
 observer an especially challenging problem. As mentioned earlier, \cite{Luis} and \cite{XZ} both focus on this problem
 but neither paper  presents a constructive solution.

 One of the main contributions of this paper is to explain why despite  this challenge,
 it is possible to correctly design the type of closed-loop observer needed  based on the
  open-loop estimator  proposed in
  \cite{park2016design,wang2017distributed}.
The construction works no matter how the $F_i$'s are defined  and is essentially the same as
   the construction  in the open-loop case  when there are no terms of the form
$\sum_{j=1}^{m} B_jF_j(e_i - e_j) $ to contend with. Moreover the construction  for the discrete-time case
involves exactly the same steps as the continuous-time case. It is unlikely that  the observers discussed in
\cite{kim2016distributed,han2018simple,wang2019distributed-cont} can be modified for this purpose because
certain key structural properties upon which the open-loop versions of these observers depend are lost when the signals
 $\sum_{j=1}^{m} B_jF_j(e_i - e_j) $ are present.

\vspace{-0.4cm}

\subsection{Distributed State Estimation for a Process Without Feedback}\label{pop}

\vspace{-0.1cm}

 A variety of distributed  estimators have been proposed in the literature for estimating the state
   of \rep{mcs} or \rep{mcsd} assuming $u_i = 0$, $i \in \mathbf{m}$ \cite{khan2010connectivity, park2016design,
   wang2017distributed,
 mitra2018distributed, kim2016distributed, han2018simple, wang2019distributed-cont, wang2019distributed-disc,
  wang2019hybrid, li2017robust}. Among these, so far only
    the distributed observers discussed in \cite{park2016design,wang2017distributed},
    seem to be amenable to application in a
    feedback loop. For this reason,
    the distributed observer studied in \cite{ wang2017distributed} will be used in this paper.
Assuming $\mathbb{N}$ is strongly connected, this particular observer is
 described by the equations
\begin{align}
\dot{x}_i &= (A \hspace{-0.3mm} + \hspace{-0.3mm} K_iC_i)x_i \hspace{-0.3mm} - \hspace{-0.3mm} K_iy_i \hspace{-0.3mm}
+ \hspace{-0.3mm} \sum_{j \in \scr{N}_i} \hspace{-0.8mm} H_{ij}(x_i-x_j) \hspace{-0.3mm} + \hspace{-0.3mm} \delta_{iq}
 \bar{C}z \label{mm1} \\
\dot{z} &= \bar{A}z + \bar{K} C_q x_q - \bar{K} y_q + \sum_{j \in \scr{N}_q} \bar{H}_{j} (x_q-x_j) \label{mm2}
\end{align}
where for $i \in \mathbf{m}$, $x_i \in \R^n$ is the estimated state, $z \in \R^{m-1}$ is the state of channel controller (see below) which is the part of agent $q$'s controller not shared with other agents, and the $K_i$, $H_{ij}$, $\bar{A}$, $\bar{K}$, $\bar{H}_j$, $\bar{C}$ are
matrices of appropriate sizes. Here, $\delta\{\cdot\}$ is the Kronecker delta and
$q$ is any pre-selected integer in $\mathbf{m}$. 
The subsystem consisting of \rep{mm2} and the signal $\delta_{iq}\bar{C}z$ is
 called a {\em channel controller} of \rep{mm1}. Its function will be explained in a moment.

The error system for this observer is described by the equations
\begin{eqnarray}
\dot{e}_i \!\!\!\! & = & \!\!\!\! (A + K_iC_i)e_i + \sum_{j \in \scr{N}_i} H_{ij}(e_i - e_j)
 + \delta_{iq} \bar{C} z \label{xe1}\\
\dot{z} \!\!\!\! & = & \!\!\!\! \bar{A}z + \bar{K} C_q e_q + \sum_{j \in \scr{N}_q} \bar{H}_{j} (e_q - e_j) \label{xe2}
\end{eqnarray}
where for each $i \in \mathbf{m}$, $e_i$ is the $i$th state estimation error $e_i= x_i-x$. Note that
\rep{xe1} and \rep{xe2} together form an $(mn+m-1)$-dimensional, unforced linear system.
 It is known that if $\mathbb{N}$ is strongly connected,  this {\em error system}'s spectrum can be freely assigned
 by appropriately picking the matrices $K_i$, $H_{ij}$, $\bar{A}$, $\bar{K}$, $\bar{H}_j$, and $\bar{C}$
  \cite{wang2017distributed}. Thus by so choosing these matrices, all of the $e_i$ and $z$ can
   be made to converge to zero exponentially fast at a pre-assigned rate.

There are several steps involved in picking the matrices
 $K_i$, $H_{ij}$, $\bar{A}$, $\bar{K}$, $\bar{H}_j$, and $\bar{C}$. First $q$ is chosen; any value of $q \in \mathbf{m}$ suffices. The next step is to temporarily ignore the channel controller \rep{xe2} and to choose matrices $\tilde{K}_i$ and the $\tilde{H}_{ij}$ so that the {\em open-loop error system}
\eq{
\dot{e}_i = (A + \tilde{K}_i C_i) e_i + \sum_{j \in \scr{N}_i} \tilde{H}_{ij} (e_i - e_j) + \delta_{iq} \tilde{u}_q
\label{ppe1}
}
$i \in \mathbf{m}$, is controllable by $\tilde{u}_q$ and observable through
\begin{equation} \label{ppe2}
\tilde{y}_q =
\begin{bmatrix}
C_q e_q \\
e_q - e_{j_1^q} \\
e_q - e_{j_2^q} \\
\vdots \\
e_q - e_{j_{m_q}^q}
\end{bmatrix}
\end{equation}
where $\{j_1^q$, $j_2^q$, $\dots$, $j_{m_q}^q\} = \scr{N}_q$. In fact, the set of $\tilde{K}_i$ and $\tilde{H}_{ij}$, $j \in \scr{N}_i$, for which these properties hold is the complement of a proper algebraic set in the linear space of all such matrices \cite{wang2017distributed}. Thus almost any choice for these matrices will accomplish the desired objective.

The next step is to pick matrices $\bar{A}$, $\bar{B}$, $\bar{C}$ and $\bar{D}$ so that the
 closed-loop spectrum of the system consisting of \rep{ppe1}, \rep{ppe2}, and the channel controller
$$
\tilde{u}_q = \bar{C}z + \bar{D} \tilde{y}_q, \;\;\;\;\; \dot{z} = \bar{A}z + \bar{B} \tilde{y}_q
$$
has the prescribed spectrum. One technique for choosing these matrices can be found in \cite{brasch1970pole}.
 Of course since the system defined by \rep{ppe1} and \rep{ppe2} is controllable and observable,
  there are many ways to define a channel controller and thus the matrices $\bar{A}$, $\bar{B}$, $\bar{C}$,
   and $\bar{D}$. In any event, once these matrices are chosen, the $K_i$ and $H_{ij}$ are defined so that
   for all $i \neq q$, $K_i \dfb \tilde{K}_i$ and $H_{ij} \dfb \tilde{H}_{ij}$, $j \in \scr{N}_i$; while
    for $i = q$, $K_q \dfb \tilde{K}_q + \hat{K}_q$ and $H_{qj} \dfb \tilde{H}_{qj} + \hat{H}_{qj}$, $j
    \in \scr{N}_q$, where $\big[\begin{matrix} \hat{K}_q & \hat{H}_{qj_1^q} & \cdots & \hat{H}_{qj_{m_q}^q} \end{matrix}\big] = \bar{D}$.
     Finally $\bar{K}$ and the $\bar{H}_{j}$, $j \in \scr{N}_q$, are defined so that $\big[\begin{matrix} \bar{K} &
      \bar{H}_{j_1^q} & \cdots & \bar{H}_{j_{m_q}^q} \end{matrix}\big] = \bar{B}$.

\vspace{-0.3cm}

\subsection{Distributed State Estimation for a Process With Feedback}\label{inputs}

In analogy with the centralized case, the first step in the development of a distributed observer-based feedback system
 is to devise state feedback laws $u_i = F_i x$, $i \in \mathbf{m}$, which endow the closed-loop system
 $\dot{x} = \left(A +\sum_{i=1}^m B_i F_i\right)x$ with prescribed properties such as stability
  and/or optimality with respect to some performance index. This is essentially a centralized design problem  since for any $F$ of appropriate size it
 is easy to construct matrices
$F_i$ such that $BF =\sum_{i=1}^m B_i F_i$ where $B = \text{row}\{B_1,B_2,\ldots, B_m\}$.
Thus centralized control techniques such as spectrum assignment
 can be used to construct $F_i$ so that the convergence
   of the state transition matrix of $A+\sum_{i=1}^m B_i F_i$  to zero is as fast as desired.

Having chosen the $F_i$,
 the next step is to   invoke certainty equivalence by replacing
 the controls $u_i = F_i x$, $i\in\mathbf{m}$, with the controls $u_i = F_i x_i$, $i\in\mathbf{m} $, where each $x_i$
is an estimate of $x$ generated by the observer discussed in \S \ref{pop},
modified to take into account the  control inputs
$u_i$, $i\in\mathbf{m}$. Doing this results in the system
\eq{
\dot{x} = Ax + \sum_{i\in\mathbf{m}}B_iF_i x_i \label{2ssxmcs}}

Modifying the open-loop distributed observer described in \S \ref{pop}  in accordance with
the substitution rule, results  in the distributed  observer described by
\begin{eqnarray}\dot{x}_i \!\!\!\!\! &=& \!\!\!\!\! (A + K_iC_i)x_i - K_iy_i + \sum_{j \in \scr{N}_i} H_{ij}(x_i-x_j) \nonumber \\
&\ & \hspace{5mm} + \hspace{1mm} \delta_{iq}\bar{C}z + \left (\sum_{j=1}^m B_j F_j\right )x_i, \;\;\; i \in \mathbf{m} \label{me1} \\
\dot{z} \!\!\!\!\! &=& \!\!\!\!\! \bar{A}z  + \bar{K}C_qx_q - \bar{K} y_q + \sum_{j \in \scr{N}_q} \bar{H}_{j} (x_q-x_j) \label{me2}
\end{eqnarray}
This is the observer which will be considered.
In the sequel it will be shown that even with this modification,
 this distributed observer  can still provide the required estimates of $x$.

Note that  the error system for \rep{me1} and \rep{me2} is described by the  linear system
\begin{eqnarray}
\dot{e}_i \!\!\!\! &=& \!\!\!\! (A+K_iC_i)e_i + \sum_{j \in \scr{N}_i} H_{ij}(e_i-e_j) + \delta_{iq}\bar{C}z \nonumber \\
& \ & \hspace{10mm} + \hspace{1mm} \sum_{j=1}^m B_j F_j (e_i-e_j), \;\;\; i \in \mathbf{m} \label{eme1} \\
\dot{z} \!\!\!\! &=& \!\!\!\! \bar{A}z + \bar{K} C_q e_q + \sum_{j \in \scr{N}_q} \bar{H}_{j}(e_q-e_j) \label{eme2}
\end{eqnarray}
Since \rep{eme1} and \rep{eme2} comprise an unforced linear system, the convergence of the $e_i$ and $z$ to zero  is determined primarily
by the error system's  spectrum, just like in the input-free case. Note in addition, that
 the process dynamics modeled by \rep{2ssxmcs} can be rewritten as
\begin{equation*} \label{3ssxmcs}
\dot{x} = \left (A +\sum_{i=1}^m B_iF_i\right )x + \sum_{i=1}^m B_iF_ie_i
\end{equation*}
Thus if the spectrum of the error system can be assigned freely, then clearly the  state of the closed-loop system determined by \rep{2ssxmcs},
\rep{me1}, and \rep{me2} can be made to converge to zero with any prescribed rate. We now explain how to assign  the  error system's spectrum.


  In the sequel, it will be
 explained how to choose the $K_i$,  $H_{ij}$, $\bar{K}$, $\bar{H}_j$, $\bar{A}$, and $\bar{C}$
  for any given $F_i$, so that  the spectrum
  of \rep{eme1} and \rep{eme2} coincides
  with a prescribed symmetric set of complex numbers, assuming $\mathbb{N}$  is strongly connected. To achieve this, attention will first be focused on the properties
  of the open-loop error system described by
\begin{multline} \label{auditt}
\dot{e}_i = (A+K_iC_i)e_i + \sum_{j \in \scr{N}_i} H_{ij}(e_i-e_j) \\
+ \sum_{j=1}^m B_jF_j(e_i-e_j) + \delta_{iq} \tilde{u}_q, \;\;\; i \in \mathbf{m}
\end{multline}
and \rep{ppe2}. This system is what results when the channel controller appearing in \rep{eme2} is removed.
 The main technical result of this is as follows.

\begin{proposition} \label{maint}
Suppose $\mathbb{N}$ is strongly connected. There are matrices
 $K_i$ and $H_{ij}$, $i \in \mathbf{m}$, $j\in\scr{N}_i$, such that for all $q \in \mathbf{m}$, the open-loop error system described by \rep{auditt} and \rep{ppe2} is observable through $\tilde{y}_q$ and controllable by $\tilde{u}_q$ with controllability index $m$.
\end{proposition}

The implication of this proposition is clear. It is possible to choose the coefficient matrices $\bar{K}$, $\bar{H}_j$, $\bar{A}$, and $\bar{C}$  which define the
channel controller to freely assign the  closed-loop spectrum of \rep{eme1} and \rep{eme2}. We are led to the following result.

\begin{theorem} \label{thm:obs-main} Suppose $\mathbb{N}$ is strongly connected.
For any set of feedback matrices $F_i$, $i \in \mathbf{m}$, any integer $q \in \mathbf{m}$, and any symmetric set $\Lambda$
of $mn + m-1$ complex numbers, there are matrices $K_i$,  $H_{ij}$, $\bar{K}$, $\bar{H}_j$, $\bar{A}$, and $\bar{C}$ for
which the spectrum of the closed-loop error system defined by \rep{eme1} and \rep{eme2} is $\Lambda$.
\end{theorem}

\begin{remark} \label{rmk:central-design}
Before proceeding, it is worth noting that while the
 local agent  controllers can be
  implemented in a distributed manner, they  collectively require a centralized design in
   which all coefficient matrices are constructed
  using a single centralized design procedure.
 Centralized designs are implicitly assumed in
   almost all decentralized control and distributed control algorithms involving
    feedback, such as the work in \cite{wang1973stabilization, corfmat1976decentralized} as well
    as the work referred to earlier in \cite{Luis,plug}.
Preliminary  efforts have also been made in \cite{Luis,plug} to develop distributed controls relying on distributed designs.
    Whether or not
     the  inevitable complexities and operational limitations introduced  by these generalization can
     lead to practical  feedback algorithms which exhibit at least  some degree of
      noise tolerance and robustness to unmodeled dynamics,
     remains to be seen. For sure,  in situations where centralized design is acceptable,
      it is difficult to imagine a situation where one would want to opt for a distributed design.
 Of course there are some distributed algorithms such as those studied in \cite{fullmer2018distributed, mou2015distributed}
 which do not call for centralized designs, but they are not feedback control algorithms.
\end{remark}

\vspace{-0.3cm}

\subsection{Analysis} \label{analysis}
We will now proceed to justify Proposition \ref{maint}.
Towards this end, note that by introducing
 the joint estimation error $\epsilon = \text{column}\, \{e_1, e_2, \dots, e_m\}$,
the open-loop error system dynamics \rep{auditt}  can be written  compactly as
\eq{
\dot{\epsilon} = \left (\tilde {A}  +
\sum_{i=1}^m \tilde{B}_i(K_i\hat{C}_i + H_{i}\tilde{C}_{i}) \right ) \epsilon + \tilde{B}_q \tilde{u}_q \label{com.o}
}
where
$\tilde{A} = I_{m \times m} \otimes( A+ \sum_{j=1}^m B_jF_j) - Q$ and for $i\in\mathbf{m}$,
 $\tilde{B}_i = b_i \otimes I_{n \times n}$,  $b_i$ is the $i$th unit vector in $\R^m$,
   $\hat{C}_i = C_i\tilde{B}_i'$  and $\otimes $ is the Kronecker product.
Here $Q$ is the $nm \times nm$
 block partitioned matrix of $m^2$ square blocks whose $ij$th block is $B_jF_j$,
 $H_i = \big[\begin{matrix} H_{ij_1^i} & H_{ij_2^i} & \cdots & H_{ij_{m_i}^i} \end{matrix}\big]$
   where $\{j_1^i$, $j_2^i$, $\dots$, $j_{m_i}^i\} = \scr{N}_i$, $\tilde{C}_i =
    \text{column} \{C_{ij_1^i}$, $C_{ij_2^i}$, $\dots$, $C_{ij_{m_i}^i}\}$
     where $C_{ij} = c_{ij} \otimes I_{n \times n}$, $j \in \scr{N}_i$, $i \in \mathbf{m}$, and $c_{ij} = b_i' - b_j'$.

Next observe that \rep{com.o} is what results in when the distributed feedback law $\tilde{v}_i = \big[\begin{matrix} K_i & H_i \end{matrix}\big] \tilde{y}_i + \delta_{iq}\tilde{u}_q$, $i \in \mathbf{m}$, is applied to the $m$-channel linear system
\eq{
\dot{\epsilon} = \tilde{A}\epsilon + \sum_{j=1}^m \tilde{B}_j \tilde{v}_j,
\;\;\; \tilde{y}_i = \begin{bmatrix} \hat{C}_i \\ \tilde{C}_i \end{bmatrix} \epsilon, \;\;\; i \in \mathbf{m} \label{anal}
}
The proof of Proposition \ref{maint} depends on the following lemmas.

\begin{lemma} \label{lanal}
The $m$-channel linear system described by \rep{anal} is jointly controllable and jointly observable.
\end{lemma}

\noindent{\bf Proof of Lemma \ref{lanal}:}
In view of the definitions of the $\tilde{B}_i$, it is clear that $\big[\begin{matrix} \tilde{B}_1 & \tilde{B}_2 & \cdots & \tilde{B}_m \end{matrix}\big]$ is the $nm \times nm$ identity. Therefore, \rep{anal} is jointly controllable.

To establish joint observability, suppose that $\tilde{v}$ is an eigenvector
of $\tilde{A}$ for which $[\hat{C}_i'\; \tilde{C}_i']' \tilde{v} = 0$, $i \in
\mathbf{m}$. From the relations $\tilde{C}_i \tilde{v} = 0$, the
 definitions of the $\tilde{C}_i$, and the assumption that $\mathbb{N}$ is strongly connected,
 it follows that $\tilde{v} = \text{column} \{v$, $v$, $\dots$, $v\}$ for some vector $v \in \R^n$.
  Meanwhile from the relations $\hat{C}_i\tilde{v} = 0$, $i \in \mathbf{m}$, and the definitions of
   the $\hat{C}_i$, it follows that $C_iv = 0$, $i \in \mathbf{m}$. Moreover, from the definition
   of $\tilde{A}$ and the structure of $\tilde{v}$, it is clear that $\tilde{A}\tilde{v} =
    ( I_{m \times m} \otimes A) \tilde{v} = \text{column} \{Av$, $Av$, $\dots$, $Av\}$. This and
     the hypothesis that $\tilde{v}$ is an eigenvector of $\tilde{A}$ imply that $v$ must be an eigenvector
      of $A$. But this is impossible because of joint observability of \rep{mcs} and the fact that $C_iv = 0$,
      $i \in \mathbf{m}$. Thus \rep{anal} has no unobservable modes through the combined outputs $\tilde{y}_i$,
      $i \in \mathbf{m}$, which means that the system is jointly observable. \hfill \qed

\begin{lemma} \label{daniel}
Suppose $\mathbb{N}$ is strongly connected. There are matrices $\hat{H}_i$, $i \in \mathbf{m}$, for
which $\left ( \sum_{i=1}^m \tilde{B}_i\hat{H}_{i}\tilde{C}_{i}, \tilde{B}_q\right )$ is a controllable pair with controllability index $m$ for every choice of $q \in \mathbf{m}$.
\end{lemma}

\noindent{\bf Proof of Lemma \ref{daniel}:} With $\bar{\scr{N}}_i$ denoting the complement of $\{i\}$ in $\scr{N}_i$,
it is shown in the proof of Proposition 1 in \cite{wang2017distributed} that there are scalars\footnote{In
\cite{wang2017distributed}
the  symbols $f_{ij}$  and $p$ are used
instead of  the symbols $\phi_{ij}$ and $q$ respectively used here.}
 $\phi_{ij}$
 for which  the matrix pair
$\left (\sum_{i\in\mathbf{m}}\sum_{j\in\bar{\scr{N}_i}} b_i\phi_{ij}c_{ij},\; b_q\right )$
is controllable for all $q\in\mathbf{m}$. Fix any such  $\phi_{ij}$  and pick any set of scalars
 $\phi_{ii},i\in\mathbf{m}$.   Next define
$D = \sum_{i\in\mathbf{m}}\sum_{j\in\scr{N}_i} b_i\phi_{ij}c_{ij}$ and  note that
$D = \sum_{i\in\mathbf{m}}\sum_{j\in\bar{\scr{N}_i}} b_i\phi_{ij}c_{ij}$ because $c_{ii} = 0,\;i\in\mathbf{m}$.
Therefore $(D,b_q)$ is also a controllable pair for each $q\in\mathbf{m}$. 

Define
$\hat{H_i} = \big[\begin{matrix} \phi_{ij_1^i} & \phi_{ij_2^i} & \cdots & \phi_{ij_{m_i}^i} \end{matrix}\big] \otimes I_n$, where as before
 $\{j_1^i$, $j_2^i$, $\dots$, $j_{m_i}^i\} = \scr{N}_i$. 
 Note that for each $i\in\mathbf{m}$, $\tilde{B}_i\hat{H}_i\tilde{C}_i = \left(
\sum_{j\in\scr{N}_i} b_i\phi_{ij}c_{ij}\right)\otimes I_n$.
Thus the matrix
$\tilde{D} \dfb  \sum_{i=1}^m \tilde{B}_i\hat{H}_{i}\tilde{C}_{i} = D\otimes I_n$.
Hence for all $q\in\mathbf{m}$,
$$
\left[
\tilde{B}_q \enspace \tilde{D}\tilde{B}_q \enspace \ldots \enspace
 \tilde{D}^{m-1}\tilde{B}_q
\right] =
\left[
b_q \enspace D b_q \enspace \cdots \enspace D^{m-1}b_q
\right] \otimes I_n.
$$
Since each pair  $(D,b_q)$ is controllable,
$$\rank \Big[\begin{matrix} \tilde{B}_q & \tilde{D} \tilde{B}_q & \cdots &
 \tilde{D}^{m-1}\tilde{B}_q \end{matrix}\Big] = mn.$$
Therefore, for each $q\in\mathbf{m}$,
$(\tilde{D},\tilde{B}_q)$ is a controllable pair with controllability index $m$. \hfill $\qed$ 



\begin{lemma} \label{g}
Let $(A_{n\times n}, B_{n\times r})$ be a real-valued controllable matrix pair with controllability index $m$. For each real matrix $M_{n\times n}$, the matrix pair $(M+gA, B)$ is controllable with controllability index no greater than $m$ for all but at most a finite number of values of the real scalar gain $g$. Moreover if $mr=n$, then $m$ is the controllability index of $(M+gA, B)$ for all but a finite number of values of $g$.
\end{lemma}

\noindent{\bf Proof of Lemma \ref{g}:}
The assumed properties of the pair $(A,B)$ imply that $mr \geq n$ and that
 there must be a minor of order $n$ of the matrix $\big[\begin{matrix} B & AB & \cdots
 & A^{m-1}B \end{matrix}\big]$ which is nonzero. Let $1$, $2$, $\dots$, $p$ be a labeling of
  the $n$th order minors of $\big[\begin{matrix} B &AB &\cdots &A^{m-1}B  \end{matrix}\big]$ and suppose that
   the $k$th such minor is nonzero. Let $\mu:\R^{n\times n} \oplus \R^{n\times r}
    \rightarrow \R$ denote the function which assigns to any matrix pair $(\bar{A}_{n\times n},
    \bar{B}_{n\times r})$ the value of the $k$th minor of
    $\big[\begin{matrix} \bar{B} &\bar{A}\bar{B} &\cdots &\bar{A}^{m-1}\bar{B} \end{matrix}\big]$.
    Thus $\mu(A,B) \neq 0$ and if $(\bar{A}, \bar{B})$ is a
     matrix pair for which $\mu(\bar{A}, \bar{B})\neq 0$, then $(\bar{A},\bar{B})$ is a
      controllable pair with controllability index no greater than $m$.

Note that $\mu(\lambda M+ A, B)$ is a polynomial in the scalar variable $\lambda$.
Since $\mu(\lambda M+ A, B)|_{\lambda = 0} \neq 0$, $\mu(\lambda M+ A, B)$ is
 not the zero polynomial. It follows that there are at most a finite number of values of $\lambda$
 for which $\mu(\lambda M+ A,B)$ vanishes and $\lambda =0$ is not one of them. Let $g$ be any number
  for which $\mu(\frac{1}{g}M+A, B) \neq 0$.
Since
 $\mu(M+gA, gB) = g^j\mu(\frac{1}{g}M+A, B)$ for some integer $j\geq 0$, it must be true that
 $\mu(M+gA, gB) \neq 0$ and thus 
   $\mu(M+gA,B) \neq 0$.
   Therefore $(M+gA, B)$ is a controllable pair with controllability index no greater than $m$.

Let $m_g$ denote the controllability index of $(M+gA, B)$; then $m_g r \geq n$.
 Suppose that $mr = n$. It follows that $m_g r \geq mr$ and thus that $m_g \geq m$.
 But for all but at most a finite set of values of $g$, $m_g \leq m$. Therefore, $m_g=m$ for
 all but at most a finite set of values of $g$. \hfill \qed

\begin{lemma} \label{lem2} Suppose $\mathbb{N}$ is strongly connected.
For any given set of appropriately sized matrices $K_i$, $i \in \mathbf{m}$, there exist matrices $H_i$ for which the matrix pair 
$$
\left( \tilde {A} + \sum_{i=1}^m \tilde{B}_i(K_i\hat{C}_i+H_{i}\tilde{C}_{i}), \tilde{B}_q \right)
$$ 
is controllable with controllability index $m$ for every $q \in \mathbf{m}$.
\end{lemma}

\noindent{\bf Proof of Lemma \ref{lem2}:}
As an immediate consequence of Lemma \ref{daniel} and Lemma \ref{g},
it is clear that for any given $K_i$, $i \in \mathbf{m}$, and for all
 but a finite number of values of $g$, the
 matrix pair $\left (\tilde {A} + \sum_{i=1}^m \tilde{B}_i(K_i\hat{C}_i + g\hat{H}_{i}\tilde{C}_{i}), \tilde{B}_q\right )$ is
  controllable with controllability index $m$ for every $q \in \mathbf{m}$. Setting $H_i = g\hat{H}_i$ thus gives the desired result. \hfill \qed

\noindent{\bf Proof of Proposition \ref{maint}:} Let $\scr{L}$ denote the collection of all lists $\{(K_i,H_i),\;i\in\mathbf{m}\}$, where for each $i$,
$(K_i,H_i)$ is a pair of matrices  sized so that  the products $\tilde{B}_iK_i\hat{C}_i $ and $\tilde{B}_iH_i\tilde{C}_i$ are defined.
 Then $\scr{L}$ is a linear space.
Let $\scr{C}$ denote the subset of lists in $\scr{L}$ for which $\left (\tilde {A} + \sum_{i=1}^m
\tilde{B}_i(K_i\hat{C}_i + H_{i}\tilde{C}_{i}), \tilde{B}_q\right )$ is
  controllable with controllability index $m$ for every $q \in \mathbf{m}$. In view of
  Lemma \ref{lem2}, $\scr{C}$ is nonempty. Moreover, since the complement of $\scr{C}$ in $\scr{L}$ coincides with  the set of solutions to a finite set of
  algebraic  equations, $\scr{C}$ is the complement of a proper algebraic set in $\scr{L}$.

By Lemma \ref{lanal}, \rep{anal} is jointly controllable. Moreover $\scr{C}$ is nonempty. Therefore, by
Theorem 1 of \cite{corfmat1976decentralized}, each complementary subsystem of
 \rep{anal} is complete.

 By Lemma \ref{lanal}, \rep{anal} is also jointly observable.
 But \rep{anal} is also complete. Therefore by Corollary 1 of \cite{corfmat1976decentralized}, the set of lists $\scr{O}\subset \scr{L}$ for which
 \rep{anal} is observable through $\tilde{y}_q$ for all $q\in\mathbf{m}$, is nonempty and thus the complement of a proper algebraic set in $\scr{L}$.
 Since the  union of proper algebraic sets in
  a linear space is also proper, $\scr{C}\cap\scr{O}$ is also the complement of a proper algebraic set in $\scr{L}$. Therefore any list in
the nonempty set $\scr{C}\cap\scr{O}$ has the required property. \hfill $\qed $

\noindent{\bf Construction:} The steps involved in constructing the matrices $F_i,K_i,H_{ij}, \bar{A},\bar{K},\bar{H}_j,$ and $\bar{C}$ which define the observer-based 
 distributed  control of interest, can be summarized as follows.
  First   state feedback matrices $F_i,\;i\in\mathbf{m}$  are chosen as explained at the beginning of \S\ref{inputs} to ensure, among other things, that the state transition matrix of  $A+\sum_{i\in\mathbf{m}}B_iF_i$ has a prescribed convergence rate. The  
  remaining matrices $K_i$, $H_{ij}$, $\bar{A}$, $\bar{K}$, $\bar{H}_j$, and $\bar{C}$ are then chosen  
  to endow the autonomous error system \rep{eme1} , \rep{eme2} with a prescribed convergence rate; this is done   in exactly the same way as these matrices  would have been  chosen  \{as is  explained at the end of \S \ref{pop}\}, had the 
   objective been 
  to provide
  the error system \rep{xe1}, \rep{xe2} with the same convergence rate -- except of course, in the current situation one would use $A+\sum_{i\in\mathbf{m}}B_iF_i$ in the calculations  rather than just $A$.

\vspace{-0.3cm}

\section{Distributed Set-Point Control} \label{sec:set-point}
Although the  substitution rule  discussed in \S \ref{sec:dis-obs-contr} can handle candidate control inputs of the forms $u_i=F_ix$, $i\in\mathbf{m}$,
 the rule is of no help in dealing with
  exogenous  reference signals $r_i$    which might appear  as additive terms in the candidate control signals; i.e.,
     $u_i = F_ix +r_i$, $i\in\mathbf{m}$.  Nonetheless certain classes of
  exogenous reference signals such as steps, ramps, and  sinusoids can in fact be dealt with without violating distributional constraints,
  if the goal is signal tracking.
The aim of this section is to illustrate this by explaining  how  the ideas discussed in the preceding section can be used to
 solve the ``distributed set-point control problem''.
 This problem will be formulated assuming that each agent $i$
 senses a scalar output $y_i = c_ix$ with the goal of adjusting
  $y_i$ to a prescribed number $r_i$ which is agent $i$'s desired
  {\em set-point} value. The {\em distributed set-point control problem} is
   then to develop a distributed feedback control system for a process modeled by
    the multi-channel system \rep{mcs} which, when applied, will enable each and every
     agent to independently adjust its output to any desired set-point value.

To construct such a control system, each agent $i$ will make use of integrator dynamics of the form
\eq{
\dot{w}_i = y_i-r_i, \;\;\; i \in \mathbf{m} \label{int}
}
where $r_i$ is the desired (constant) value to which $y_i$ is to be set. The combination
 of these integrator equations plus the multi-channel system described by \rep{mcs}, is
 thus a system of the form
\eq{
\dot{\tilde{x}} = \tilde{A}\tilde{x} + \sum_{i=1}^m \tilde{B}_i u_i - \tilde r, \hspace{0.3in} w_i =
 \tilde{c}_i\tilde{x}, \;\;\; i \in \mathbf{m} \label{b1}
}
where $\tilde{x} = \text{column} \{x, w_1, w_2, \dots, w_m\}$,
$$
\tilde{A} = \begin{bmatrix} A & \mathbf{0} \\ C & \mathbf{0} \end{bmatrix}, \;\;\;\;\;\; \tilde{B}_i =
\begin{bmatrix} B_i \cr \mathbf{0} \end{bmatrix},\; i \in \mathbf{m}, \;\;\;\;\; \tilde{r} = \begin{bmatrix}\mathbf{0} \cr r \end{bmatrix}
$$
$C = \text{column} \{c_1$, $c_2$, $\dots$, $c_m\}_{m \times n}$, $r = \text{column} \{r_1$,
$r_2$, $\dots$, $r_m\}$, and $\tilde{c}_i = \big[\begin{matrix} \mathbf{0} & e_i' \end{matrix}\big]$, $e_i$ being the $i$th unit vector
in $\R^m$. Thus \rep{b1} is an $(n+m)$-dimensional, $m$-channel system with measurable outputs $w_i$, $i
\in \mathbf{m}$, control inputs $u_i$, $i \in \mathbf{m}$, and constant exogenous input $\tilde{r}$. Note
that {\em any} linear time-invariant feedback control, distributed or not, which stabilizes this system,
will enable each agent to attain its desired set-point value. The reason for this is simple. First note
that any such control will bound the state of the resulting closed-loop system and cause the state to tend
to a constant limit as $t \rightarrow \infty$. Therefore, since each $w_i$ is a state variable, each must
 tend to a finite limit. Similarly, $x$ and thus each $y_i$ must also tend to a finite limit. In view of
  \rep{int}, the only way this can happen is if each $y_i$ tends to agent $i$'s desired set-point value $r_i$.

To solve the distributed set-point control problem, it is enough to devise a distributed controller which stabilizes \rep{b1}.
 This can be accomplished using the ideas discussed earlier in \S \ref{sec:dis-obs-contr}, provided that \rep{b1} is both jointly controllable by the $u_i$ and jointly observable through the $w_i$. According to Hautus's lemma \cite{hautus1969controllability}, the condition for joint observability is that
$$
\rank \begin{bmatrix} sI- \tilde{A} \\ \tilde{C} \end{bmatrix} = n+m
$$
for all complex number $s$, where $\tilde{C} = \text{column} \{\tilde{c}_1$, $\tilde{c}_2$, $\dots$, $\tilde{c}_m\}$. In other words, what is required is that
\eq{
\rank \begin{bmatrix} sI_n - A & \mathbf{0} \\ -C & sI_m \\ \mathbf{0} & I_m \end{bmatrix} = n+m \label{ober}
}
But $(C, A)$ is an observable pair because \rep{mcs} is a jointly observable system. From this, the Hautus condition, and the structure of the matrix pencil appearing in \rep{ober}, it is clear that the required rank condition is satisfied and thus that \rep{b1} is a jointly observable system.

To establish joint controllability of \rep{b1}, it is enough to show that $\rank \big[\begin{matrix} sI-\tilde{A} & \tilde{B} \end{matrix}\big] = n+m$ for all complex number $s$, where $\tilde{B} = \big[\begin{matrix} \tilde{B}_1 & \tilde{B}_2 & \cdots & \tilde{B}_m \end{matrix}\big]$. In other words, what is required is that
\eq{
\rank \hspace{-1mm} \begin{bmatrix}sI_n - A \! & \! \mathbf{0} \! & \! B_1 \! & \! B_2 \! & \! \cdots \! & \! B_m \\
-C \! & \! sI_m \! & \! \mathbf{0} \! & \! \mathbf{0} \! & \! \cdots \! & \! \mathbf{0} \end{bmatrix} = n+m \label{thy}
}
But since \rep{mcs} is jointly controllable, $\rank \left[sI-A \quad B\right] = n$ for all $s$, where $B = \left[B_1 \quad B_2 \quad \cdots \quad B_m\right]$. Thus \rep{thy} holds for all $s \neq 0$. For $s=0$, \rep{thy} will also hold provided
\eq{
\rank \begin{bmatrix} A & B \\ C & \mathbf{0} \end{bmatrix} = n+m \label{gnd}
}
That is, \rep{gnd} is the condition for \rep{b1} to be jointly controllable and thus stabilizable with distributed control.

It is possible to give a simple interpretation of condition \rep{gnd} for the case when each $B_i$ is a single column. In this case the transfer matrix $C(sI-A)^{-1}B$ is square and condition \rep{gnd} is equivalent to the requirement that its determinant has no zeros at $s=0$ \cite{morse1973structural}. Note that if the transfer matrix were nonsingular but had a zero at $s=0$, this would lead to a pole-zero cancellation at $s=0$ because of the integrators.

Suppose condition \rep{gnd} is satisfied. In order to stabilize \rep{b1}, the first step would be to
construct a distributed observer-based control as outlined in \S \ref{sec:dis-obs-contr}  which stabilizes the reference-signal-free system
\begin{equation*} \label{b2}
\dot{\tilde{x}} = \tilde{A}\tilde{x} + \sum_{i=1}^m \tilde{B}_iu_i\hspace{0.3in} w_i = \tilde{c}_i\tilde{x}, \;\;\; i \in \mathbf{m}
\end{equation*}
Application of such a distributed control
to \rep{b1} would then  stabilize \rep{b1} and thus provide a solution to the distributed set-point control problem,
 despite the fact that
 the signals $x_i$ would not be asymptotically correct estimates of $\tilde{x}$.

\vspace{-0.3cm}

\section{Distributed  Control Recast as Decentralized Control}\label{liu}

The  main result of \cite{kexin} states that every jointly controllable,
jointly observable continuous-time multi-channel linear system  with an associated strongly connected  neighbor graph,
can be exponentially stabilized with  a distributed dynamic compensator. Meanwhile results of \S \ref{sec:dis-obs-contr}
make it clear that under the same conditions,
 exponentially stabilized with a prescribed convergence rate can be achieved  with a  distributed observer-based control.
These findings prompt a number of questions.
For example,  is exponential stabilization still possible
  if there is no message passing across the network? Is exponential stabilization at a prescribed convergence rate  possible
  with message passing  but without a distributed observer? If yes, what messages should be passed among neighboring agents?
Under what conditions  can distributed control be used to  freely assign  the closed-loop spectrum
of a multi-channel linear system without  the  spectral separation restriction imposed by
certainty equivalence observer-based control?
These are some of the questions to which this section is addressed.

   The first question just raised, namely ``Is exponential stabilization still possible
  if there is no message passing across the network?'' is actually a question in the area of ``decentralized control''.
Accordingly, some of the main results from classical decentralized control will be reviewed next.

\vspace{-0.3cm}

 \subsection{Decentralized Control}\label{dcontrol}
 \vspace{-0.1cm}
The basic distributed control problem  formulated in \S \ref{bdcp}, but without message passing,
is precisely the classical decentralized control problem
 studied years ago in \cite{wang1973stabilization, corfmat1976decentralized}.
  Thus the {\em classical decentralized control problem} is to find $m$ linear time-invariant  controllers, one for each channel of
   \rep{mcs}, which
  stabilize the  resulting closed-loop system with any prescribed convergence rate. Much is known about this problem.
  Perhaps the most fundamental
   is the fact that  the closed-loop spectrum  which results when   time-invariant linear controllers are applied to the channels of
    \rep{mcs},
   contains a uniquely determined
   subset of eigenvalues \{or   ``fixed modes''  \cite{wang1973stabilization}\}, called the {\em fixed spectrum} of \rep{mcs}
    which is the same for all possible  linear, time-invariant
   channel controllers which might be applied. This subset, denoted by $\Lambda_{\rm fixed}$, is defined by the formula
$$\Lambda_{\rm fixed} = \bigcap_{F_i\in\R^{p_i\times q_i},\;i\in\mathbf{m}}\sigma(A+\sum_{i=1}^m B_iF_iC_i)$$
where $\sigma(\cdot)$ denotes the spectrum of a matrix. Thus $\Lambda_{\rm fixed}$ is the set of eigenvalues of
      the matrix $A+B_1F_1C_1 +B_2F_2C_2+\cdots +B_mF_mC_m$ which don't change as the real-valued matrices
      $F_1$, $F_2$, $\dots$, $F_m$ range over all possible values.
It is shown in \cite{wang1973stabilization} that the classical decentralized control problem for \rep{mcs} is solvable if and only
 if $\Lambda_{\rm fixed}$ is an empty set.
 Identical definitions and analogous  conclusions hold for the
 discrete-time multi-channel linear system defined by \rep{mcsd}.

There is a  more explicit characterization of  the elements in  $\Lambda_{\text{fixed}}$  \cite{anderson1981algebraic}.
To explain what it is, use will be made of the following notation. For each subset $\mathbf{s} = \{i_1,i_2,\ldots, i_s\} \subset
\mathbf{m}$
whose elements are
  ordered as $i_1<i_2< \dots <i_s$, let $B_{\mathbf{s}}$ and $C_{\mathbf{s}}$ denote the matrices
\begin{equation*}
B_{\mathbf{s}} =
\left[
B_{i_1} \enspace B_{i_2} \enspace \cdots \enspace B_{i_s}
\right]
\hspace{1.8mm} \text{and} \hspace{1.8mm}
C_{\mathbf{s}} =
\begin{bmatrix}
C_{i_1}'& 
C_{i_2}'&
\ldots &
C_{i_s}'
\end{bmatrix}'
\end{equation*}
respectively. A proof of the following result can be found in \cite{anderson1981algebraic}.
\begin{proposition}   \label{prp:mtrx-pencil}
Let $A $, $B_i$, and $C_i$, $i\in\mathbf{m}$, be the coefficient matrices of the  $m$-channel linear system defined by either
 \rep{mcs} or \rep{mcsd}.
 A complex number $\lambda\in\Lambda_{\text{fixed}}$ if and only if for some  nonempty, proper
  subset\footnote{Note that the rank of the matrix pencil in \rep{ineq:rank-test} is $n$ if $\mathbf{s}$ is either empty
  or equal to $\mathbf{m}$ because of joint controllability and joint observability respectively of either \rep{mcs} or \rep{mcsd}.} $\mathbf{s} \subset \mathbf{m}$,
\begin{equation} \label{ineq:rank-test}
\rank
\begin{bmatrix}
\lambda I_n - A & B_{\mathbf{m}-\mathbf{s}} \\
C_{\mathbf{s}} & \mathbf{0}
\end{bmatrix}
< n
\end{equation}
where $\mathbf{m}-\mathbf{s}$ is the complement of $\mathbf{s}$ in $\mathbf{m}$.
\end{proposition}
Obviously the condition for \rep{mcs} or \rep{mcsd} to have no fixed
eigenvalues is that the rank of the matrix pencil in
 (\ref{ineq:rank-test}) is greater than or equal to $n$ for all
  $\lambda \in \sigma(A)$ and all proper subsets $\mathbf{s} \subset \mathbf{m}$.
Use will be made of this characterization of a multi-channel system's fixed eigenvalues  at several different places in the paper.

Also of   importance in  decentralized control is the concept of the  ``transfer graph'' of a multi-channel linear system.
By the  \emph{transfer graph} of an $m$-channel linear system \rep{mcs} or \rep{mcsd} is meant the directed graph with $m$ vertices  which has  an arc from vertex $j$ to vertex $i$ whenever the
 transfer matrix $C_i (sI - A)^{-1} B_j \neq 0$ \cite{corfmat1976decentralized}.
  It is known that if  multi-channel linear system  \rep{mcs} or \rep{mcsd} has both a  strongly connected transfer graph and no fixed
  eigenvalues \{i.e., $\Lambda_{\rm fixed}$ is an empty set\}, then  there exist matrices $F_i$, $i \in \mathbf{m}$, of
  appropriate sizes such that the closed-loop system which results under application of the decentralized
   feedback laws $u_i = F_iy_i+v_i$, $i\in\mathbf{m}$, is controllable by any one input $v_j$ and  observable through any one output
    $y_k$
      \cite{corfmat1976decentralized}. Thus with a suitably defined  feedback controller    from  $y_i$ to $v_i$ \{for any one channel
      index $i\in\mathbf{m}$\},
the resulting closed-loop spectrum of the multi-channel system under consideration   can be freely assigned; such a controller
can be crafted using standard techniques such as those found in \cite{brasch1970pole}.
 If, on the other hand,   \rep{mcs} \{or \rep{mcsd}\}
      has no fixed  eigenvalues  and its transfer graph is \emph{not} strongly connected, ``free'' spectrum assignment can only
      be achieved with certain constraints on the symmetry of subsets of the spectrum.
These constraints are that the
      closed-loop spectrum must be partitioned into $\eta$ symmetric sets of complex numbers, where $\eta$ is the number of
      strongly connected components in the system's  transfer graph, and that the cardinality of each symmetric set must equal the
      dimension of the corresponding closed-loop ``strongly connected subsystem'' \cite{corfmat1976decentralized}.

\vspace{-0.3cm}
\subsection{Transmission of Just Measured Outputs Across a Network
 Is Not Sufficient for Stabilization} \label{nope}

 The aim of this subsection is to demonstrate that stabilization of  a multi-channel linear system with distributed control
 can not necessarily be achieved  if the only signal which each agent transmits to its followers
 is its measured output.  The following example illustrates this.

\noindent{\bf Example:} Consider the $m=3$ channel, jointly controllable, jointly observable linear system \rep{mcs}
 with coefficient matrices
\begin{align*} \label{expl:origin-3chnl}
A &=
\begin{bmatrix}
1 & 0 & 0 \\
0 & 1 & 0 \\
0 & 1 & 1
\end{bmatrix}
,
B_1 =
\begin{bmatrix}
1 \\
0 \\
0
\end{bmatrix}
,
B_2 =
\begin{bmatrix}
0 \\
1 \\
0
\end{bmatrix}
,
B_3 =
\begin{bmatrix}
1 \\
0 \\
0
\end{bmatrix} \nonumber \\
C_1 &=
\begin{bmatrix}
1 & 0 & 0 \\
0 & 0 & 1
\end{bmatrix}
,
C_2 =
\begin{bmatrix}
0 & 1 & 0
\end{bmatrix}
,
C_3 =
\begin{bmatrix}
0 & 1 & 0
\end{bmatrix}
\end{align*}
and the cyclic neighbor graph; i.e., $\scr{N}_1 =\{1, 2\}$, $\scr{N}_2 =\{2, 3\}$ and  $\scr{N}_3 =\{3, 1\}$.
Suppose  each agent $i$ receives only the measured output $y_j$ of neighbor $j \in\scr{N}_i$ and nothing more.
 Then for $i\in\mathbf{m}$, the  signal
agent $i$ can input to its channel controller is the {\em augmented}
 output $\bar{y}_i = \text{column}\{y_i, y_j\}$.  Clearly this distributed control problem    is the same  as the classical
   decentralized control problem for the three-channel system
   \eq{
\dot{x} = Ax + \sum_{i=1}^m B_i u_i, \hspace{5mm} \bar{y}_i = C_{\scr{N}_i} x, \hspace{5mm} i \in \mathbf{m}
\label{mss}}
where $C_{\scr{N}_1} = \text{column} \{C_1, C_2\}$, $C_{\scr{N}_2} = \text{column}
     \{C_2, C_3\}$, and $C_{\scr{N}_3}= \text{column} \{C_3, C_1\}$.
It is easy to check that
\begin{equation*}
\rank
\begin{bmatrix}
I - A & B_1 & B_3 \\
C_{\scr{N}_2} & \mathbf{0} & \mathbf{0}
\end{bmatrix}
= 2
\end{equation*}
Thus by Proposition \ref{prp:mtrx-pencil}, \rep{mss} has a fixed eigenvalue at $\lambda =1$. Therefore, \rep{mss} cannot
be stabilized by decentralized
 control, which means that \rep{mcs} cannot be stabilized by distributed control in which the only signal  each agent $i$ receives
 from its neighbor $j$ is $y_j$.

What this example illustrates is something a little counter-intuitive, namely that
  for distributed stabilization to be possible
 it is not in general
 enough for each agent to share just its measured output with its
  followers. Meanwhile, the results of \S
  \ref{pop} clearly demonstrate that to achieve stabilization
  with distributed control it
is enough for each agent $i$ to share its controller state with all of its
 followers, at least when that state is the  state $x_i$
    of agent $i$'s local state estimator. In the sequel, we expand on this observation, while being mindful of the
     limitation the preceding example illustrates.

\vspace{-0.4cm}
\subsection{Extended System} \label{esyst}
\vspace{-0.1cm}

Examination of the observer-based agent controllers described by \rep{me1} and \rep{me2} shows that all agents
employ state vectors $x_i$ which are shared across the network and, in addition, one agent uses an additional vector $z$  which is not shared.
 Prompted by this, we now consider a configuration in which each agent's controller state
consists of two  components, namely  a sub-state   $x_i \in {\R}^{n_i}$
 which agent $i$   communicates  to its followers, and an additional
sub-state $z_i \in {\R}^{k_i}$  which is not shared across the network.
From this perspective,  agent $i$'s controller is an $(n_i+k_i)$-dimensional
linear system  with output $u_i$, state $\text{column}\{x_i,z_i\}$,
 and  inputs  consisting  of   the measured output  $y_j$ and shared sub-state  $x_j$
 of each of its neighbors $j \in \scr{N}_i$.
The state dynamics for  each $x_i$ and $z_i$  are thus of the  forms
\begin{equation} \label{bun}
\dot{x}_i = v_i, \hspace{3mm} i \in \mathbf{m}
\end{equation}
and $\dot{z}_i = \nu_i$ for $i \in \mathbf{m}$ respectively,
 where $v_i$ and $\nu_i$ are appropriately chosen linear functions of $z_i$ and the signals $y_j$, $x_j$ for all $j\in\scr{N}_i$.
It will become clear soon that the original $m$-channel system \rep{mcs} and each $x_i$'s dynamics \rep{bun}
 can be combined into a new open-loop $m$-channel system called the ``extended system'', which incorporates
 the message passing required by distributed control. Then the basic distributed control problem for
  the original system \rep{mcs} can be recast as a decentralized control problem for the extended
  system defined below. To effectively control the extended system, a decentralized dynamic output
   feedback controller may be needed at each channel. From this point of view, each $z_i$ is the state
    of the decentralized dynamic controller for channel $i$ of the extended system, that is why $z_i$ is
     not shared across the network. Note that it is possible that the decentralized output feedback
     for some channel $i$ of the extended system is static, in that case $k_i = 0$, so there will be no $z_i$ in agent $i$'s controller.

There is a convenient way to describe this configuration.
Towards this end, let $\bar{x}$ denote the {\em extended state}
$$\bar{x}  =  \text{column}\{x, x_1, \ldots, x_m\}$$ and for $i\in\mathbf{m}$, define
\begin{align*}
\bar{u}_i &= \text{column}\{u_i,v_i\} \\
\bar{y}_i &= \text{column}\{y_{i_1},y_{i_2},\ldots, y_{i_{m_i}},x_{i_1},x_{i_2},\ldots, x_{i_{m_i}}\}
\end{align*}
where $\{i_1,i_2,\ldots,i_{m_{i}}\}$  is the set of labels of agent $i$'s neighbors
 in $\scr{N}_i$ arranged in ascending order.
By the {\em extended system} determined by \rep{mcs} and \rep{bun} is meant the  $\bar{n}$-dimensional, $m$-channel
linear system
\eq{ \dot{\bar{x}} =
 \bar{A}\bar{x} +\sum_{i=1}^m \bar{B}_i\bar{u}_i \hspace{.5in}  \bar{y}_i  =  \bar{C}_i\bar{x},\;\;\;i\in\mathbf{m}\label{emcs}}

\noindent where  $\bar{n} = n + \sum_{i=1}^m n_i$,
\begin{align*}
\bar{A} &= \text{diagonal} \left\lbrace
A, \mathbf{0}_{(\bar{n}-n)\times(\bar{n}-n)}
\right\rbrace,\\
\bar{B}_i &=  \text{diagonal} \left\lbrace
B_i, E_i
\right\rbrace, \hspace{3mm} i\in\mathbf{m},\\
\bar{C}_i &= \text{diagonal} \left\lbrace
C_{\scr{N}_i}, E'_{\scr{N}_i}
\right\rbrace, \hspace{3mm} i\in\mathbf{m}
\end{align*}
Here
\begin{equation*}
E_i = \text{column} \left\lbrace
\mathbf{0}_{g_i^-\times n_i}, I_{n_i\times n_i}, \mathbf{0}_{g_i^+\times n_i}
\right\rbrace
\end{equation*}
where $g_i^- = n_1+\cdots +n_{i-1}$ and $g_i^+ = n_{i+1} + \cdots + n_m$.
It is straightforward to verify that this system is jointly controllable
  and jointly observable regardless  of the connectivity of $\mathbb{N}$.

\subsection{The Distributed Control Problem  Restated}\label{restate}

With the concept of an extended system at hand, it is possible to restate the basic distributed control problem for \rep{mcs}
 posed at the beginning of \S \ref{bdcp}
 in the following way.

\noindent{\em Basic Distributed Control Problem -- Restated}: For the $m$-channel extended system described by \rep{emcs}, develop a
systematic procedure for picking integers $n_i$, $i\in\mathbf{m}$, and constructing $m$ time-invariant  feedback controllers with
$\bar{y}_i$ and $\bar{u}_i$ the input to and output from controller $i$ respectively,
 so that the state of the resulting closed-loop system converges to zero exponentially fast at a pre-assigned rate.

It is obvious that the originally stated basic distributed control problem and its restated version are entirely equivalent.
 One virtue of the revised formulation is that it enables one to easily address the basic distributed control 
  problem without being compelled to use a
  distributed observer.
The reason for this stems from the fact that for fixed $n_i$, the restated version is mathematically a
decentralized control problem.
Thus deciding whether or not the distributed control problem for \rep{mcs} can be solved amounts to deciding what
 must be true of the $n_i$ for
the extended system \rep{emcs} to have no fixed eigenvalues.
In the sequel this will be done  assuming $\mathbb{N}$ is strongly connected.
In light of the results of \S \ref{sec:dis-obs-contr} and  Proposition \ref{prp:mtrx-pencil},
it is reasonable to expect that when the  dimensions  $n_i$ of all  local controller's  shared sub-states
 are large enough to compensate for the rank
deficiency induced by the fixed eigenvalues of (\ref{mcs}), the corresponding
 extended system (\ref{emcs}) will have no fixed eigenvalues. This is indeed the case.

\begin{theorem} \label{thm:strg-graph}
Suppose     that $\mathbb{N}$ is strongly connected and that
\begin{equation} \label{eqn:n_i}
n_i \geq n - \min_{\substack{\mathbf{s} \in \bar{\mathbf{m}} \\ \lambda \in \sigma(A)}} \rank
\begin{bmatrix}
\lambda I_n - A & B_{\mathbf{m}-\mathbf{s}} \\
C_{\scr{N}_{\mathbf{s}}} & \mathbf{0}
\end{bmatrix}
, \hspace{3mm} i \in \mathbf{m}
\end{equation}
where $\bar{\mathbf{m}}$ is the set of all  nonempty proper subsets of $\mathbf{m}$.
Then the extended system defined by \rep{emcs} has no fixed eigenvalues.
\end{theorem}

Theorem \ref{thm:strg-graph} thus gives a lower bound on
the dimensions of the local controllers' shared sub-states  needed for the extended system  \rep{emcs} to have no fixed eigenvalues.
It is easy to check that in the special case when \rep{mcs} has no fixed eigenvalues, this lower bound is zero as it should be. Another observation is that the lower bound is reached for every $i \in \mathbf{m}$ if and only if for each $i \in \mathbf{m}$, there exists a subset $\mathbf{s} \subset \mathbf{m}$ such that $\scr{N}_{\mathbf{s}} \cap (\mathbf{m} - \mathbf{s}) = \{i\}$ and
\begin{equation*}
\rank \hspace{-1mm}
\begin{bmatrix}
\lambda I_n - A & B_{\mathbf{m}-\mathbf{s}} \\
C_{\scr{N}_{\mathbf{s}}} & \mathbf{0}
\end{bmatrix}
=
\min_{\substack{\mathbf{s} \in \bar{\mathbf{m}} \\ \lambda \in \sigma(A)}} \rank \hspace{-1mm}
\begin{bmatrix}
\lambda I_n - A & B_{\mathbf{m}-\mathbf{s}} \\
C_{\scr{N}_{\mathbf{s}}} & \mathbf{0}
\end{bmatrix}
\end{equation*}
Whether or not this lower bound is tight remains to be seen.

\begin{remark} \label{rmk:no-yi} It is straightforward to verify that if the matrices  $\bar{C}_i =  \text{diagonal}
\{ C_{\scr{N}_i}, E_{\scr{N}_i}'\}$ appearing in the definition of  the extended system in \rep{emcs}, are replaced with the matrices
\begin{equation*}
\bar{C}_i^{\text{modified}}=\text{diagonal}\{ C_i, E_{\scr{N}_i}'\},
\end{equation*}
the resulting modified extended system will
still be jointly observable. Moreover, as the proof of Theorem \ref{thm:strg-graph}  below readily reveals,
if hypothesis \rep{eqn:n_i} in the statement of Theorem \ref{thm:strg-graph} is replaced with the hypothesis
\begin{equation} \label{neqn:n_i}
n_i \geq n - \min_{\substack{\mathbf{s} \in \bar{\mathbf{m}} \\ \lambda \in \sigma(A)}} \rank
\begin{bmatrix}
\lambda I_n - A & B_{\mathbf{m}-\mathbf{s}} \\
C_{\mathbf{s}} & \mathbf{0}
\end{bmatrix}
, \hspace{3mm} i \in \mathbf{m}
\end{equation}
 the modified extended system will have no fixed eigenvalues. While \rep{neqn:n_i} demands more of the $n_i$ than \rep{eqn:n_i} does, because
$$\rank \begin{bmatrix} \lambda I- A & B_{\mathbf{m}-\mathbf{s}} \cr  C_{\scr{N}_{\mathbf{s}}} & 0 \end{bmatrix}
\geq \rank \begin{bmatrix} \lambda I- A & B_{\mathbf{m}-\mathbf{s}} \cr  C_{\mathbf{s}} & 0 \end{bmatrix},$$
the virtue of this modification is that only the signals $x_i$ and not the signals $y_i$ need be transmitted across the network.
\end{remark}

\noindent{\bf Proof of Theorem \ref{thm:strg-graph}:}
Suppose  that \rep{eqn:n_i} holds.  Fix $\mathbf{s}\in\bar{\mathbf{m}}$ and $\lambda\in \sigma(A)$. Let
$$\bar{M} = \begin{bmatrix} \lambda I-\bar{A} & \bar{B}_{\mathbf{m}-\mathbf{s}} \cr
\bar{C}_{\mathbf{s}} & \mathbf{0} \end{bmatrix} \;\;\;\; \text{and} \;\;\;\; M = \begin{bmatrix} \lambda I-A & B_{\mathbf{m}-\mathbf{s}} \cr
C_{\scr{N}_{\mathbf{s}}} & \mathbf{0} \end{bmatrix}$$
In view of the structure of the coefficient matrices defining \rep{emcs},
\eq{\rank \bar{M} = \rank  M + \rank E_{\mathbf{m}-\mathbf{s}}
+
\rank E_{\mathcal{N}_{\mathbf{s}}}\label{funch}}
Clearly
\eq{\rank E_{\mathbf{m}-{\mathbf{s}}} = \sum_{i\in\mathbf{m}-\mathbf{s}}\rank E_i= \sum_{i\in\mathbf{m}-\mathbf{s}} n_i\label{dear}}
and
\eq{\rank E_{\mathcal{N}_{ \mathbf{s}}} = \sum_{i\in
 \mathbf{s}} \sum_{j\in\scr{N}_i}\rank E_{j}=\sum_{i\in\mathbf{s}} \sum_{j\in\scr{N}_i}n_{j}\label{pppp}}

By hypothesis, $\mathbb{N}$ is strongly connected. This implies that
$\scr{N}_{\mathbf{s}}\cap(\mathbf{m}-\mathbf{s})$ is nonempty
 so for
some $k\in\mathbf{m}-\mathbf{s}$, there must be a $j\in\mathbf{s}$ such that $k\in\scr{N}_j$;
clearly $k\neq j$. But
 $i\in\scr{N}_i,\;i\in\mathbf{s}$.  It follows that
$$\sum_{i\in\mathbf{s}} \sum_{j\in\scr{N}_i}n_{j} \geq \sum_{i\in\mathbf{s}}n_i +n_k$$
From this, and \rep{funch} -- \rep{pppp}
it follows that
$$\rank \bar{M} \geq \rank M +\sum_{i\in\mathbf{m}}n_i + n_k$$
But from \rep{eqn:n_i}
$$
n_k \geq n -  \rank M $$
so
$$\rank \bar{M} \geq n +\sum_{i\in\mathbf{m}}n_i = \bar{n} $$
Since this holds for all $\mathbf{s} \in \bar{\mathbf{m}}$ and all $\lambda\in\sigma(A)$, the hypothesis of Proposition
  \ref{prp:mtrx-pencil} is satisfied, so
  the extended multi-channel system \rep{emcs} has no fixed eigenvalues. \hfill $\qed$

Note that the only place in the proof of Theorem \ref{thm:strg-graph} where strong connectivity of $\mathbb{N}$ is used is to ensure that
$\scr{N}_{\mathbf{s}}\cap(\mathbf{m}-\mathbf{s})$ is a nonempty set.
It is easy to verify that $\mathbb{N}$ is strongly connected if and only if $\scr{N}_{\mathbf{s}}\cap(\mathbf{m}-\mathbf{s})$ is nonempty for all $\mathbf{s}\in\bar{\mathbf{m}}$.

The example in \S \ref{nope} and Theorem \ref{thm:strg-graph} imply that
to avoid the fixed eigenvalues, it is necessary and sufficient to transmit  shared sub-states of appropriate  dimensions
of the local controllers.

\vspace{-0.3cm}

\subsection{Free Spectrum Assignability}\label{freedom}

The results of \S \ref{sec:dis-obs-contr} make it clear that the closed-loop spectrum of a multi-channel linear
system can be freely assigned
with distributed observer-based control provided the spectrum  consists of the disjoint union of two
symmetric sets, one for the distributed observer and the other for the closed-loop  process
 under state feedback. It will now be explained how this {\em partitioning constraint}  can be avoided.

As discussed at the end of \S \ref{dcontrol} and established  in \cite{corfmat1976decentralized},
the conditions for free spectrum assignability   with decentralized control are that the multi-channel
system to be controlled must have
  a strongly connected transfer graph and no fixed eigenvalues. To make use of this result here, it is necessary to relate the
  transfer graph of the extended system \rep{emcs} to the transfer graph of the original system \rep{mcs}.
   The following proposition, which utilizes the concept of the ``union'' of two directed graphs\footnote{By the {\em union} of two directed  graphs $\mathbb{G}_1$ and $\mathbb{G}_2$ with the same vertex set $\scr{V}$ is meant the graph
$\mathbb{G}_1\cup\mathbb{G}_2$ with vertex set $\scr{V}$ whose arc set is the union of the arc sets of $\mathbb{G}_1$ and $\mathbb{G}_2$.},
 accomplishes this.

\begin{proposition} \label{prp:transfer-matrix-graph}
Suppose that $n_i > 0$ for $i \in \mathbf{m}$, that $\mathbb{N}$ and $\mathbb{G}$ are
 the neighbor and transfer graphs respectively of the original system \rep{mcs}, and that $\bar{\mathbb{G}}$ is the
 transfer graph of the extended system \rep{emcs}. Then $\mathbb{N}\cup \mathbb{G}\subset\bar{\mathbb{G}}$.
\end{proposition}

\noindent \textbf{Proof of Proposition \ref{prp:transfer-matrix-graph}:}
Suppose $(j,i)$ is an arc in $\mathbb{N}\cup\mathbb{G}$ in which case $(j,i)$ is an arc in $\mathbb{N}$ or an arc in $\mathbb{G}$.
If $(j,i)$ is an arc in $\mathbb{N}$, then $j\in\scr{N}_i$, so $E_{\mathcal{N}_i}' E_j \neq 0$. On the other hand, if
$(j,i)$ is an arc in $\mathbb{G}$, then $C_i (\lambda I - A)^{-1} B_j \neq 0$, so $C_{\scr{N}_i}(\lambda I - A)^{-1} B_j
\neq 0$ since $i\in\scr{N}_i$. In either case $\bar{C}_i (\lambda I -\bar{ A})^{-1} \bar{B}_j \neq 0$,
because of the definitions of the coefficient matrices defining \rep{emcs}. Therefore $(j,i)$ is an arc in $\bar{\mathbb{G}}$. \hfill $\qed$

The proposition implies that if the neighbor graph $\mathbb{N}$ is strongly connected, then so is the transfer graph of \rep{emcs}.
We are led to the following corollary to Theorem \ref{thm:strg-graph}.

\begin{corollary} \label{cor:free-assign}
Under the hypotheses of Theorem \ref{thm:strg-graph}, the closed-loop spectrum of \rep{mcs} can be freely assigned
with distributed control.
\end{corollary}

More specifically, under the hypotheses of Theorem \ref{thm:strg-graph}, there are output feedback laws\footnote{Almost all matrices $\bar{F}_i$, $i \in \mathbf{m}$, of appropriate sizes will do \cite{corfmat1976decentralized}.} $\bar{u}_i = \bar{F}_i \bar{y}_i$, $i \in \mathbf{m}$, such that the system
\begin{equation} \label{sys:contr-obs-thru-q}
\dot{\bar{x}} = \left( \bar{A} + \sum_{i=1}^m \bar{B}_i \bar{F}_i \bar{C}_i \right) \bar{x} + \bar{B}_q \bar{u}_q, \hspace{5mm} \bar{y}_q = \bar{C}_q \bar{x}
\end{equation}
is controllable and observable through any channel $q \in \mathbf{m}$ \cite{corfmat1976decentralized}. Suppose the minimum of the controllability index and the observability index of (\ref{sys:contr-obs-thru-q}) is $k_q+1$. Then, with suitable matrices $\breve{A}$, $\breve{B}$, $\breve{C}$, and $\breve{D}$, the closed-loop system consisting of (\ref{sys:contr-obs-thru-q}) and agent $q$'s channel controller $z_q \in \R^{k_q}$ which is not shared on the network,
\begin{equation*}
\dot{z}_q = \breve{A}z_q + \breve{B} \bar{y}_q, \hspace{6mm}
\bar{u}_q = \breve{C}z_q + \breve{D} \bar{y}_q
\end{equation*}
can be exponentially stabilized at any prescribed convergence rate using a linear time-invariant control \cite{brasch1970pole}. In this way, the total dimension of distributed controllers for the original system \rep{mcs} is $k_q + \sum_{i=1}^m n_i$.

While the results derived thus far mean that the basic distributed control problem can be addressed with
 the tools from classical decentralized control theory, other
 yet to be developed design techniques  may possibly also be applied.

\vspace{-0.3cm}

\section{Transmission Delays} \label{sec:delays}

\vspace{-0.1cm}

The aim of this section is to  broaden the distributed feedback problem to account
 for possible transmission delays across the network.
To avoid the technical challenges associated with delay differential equations, the
investigation is restricted exclusively
to discrete-time systems.

 In \S \ref{subsec:conv-rate}, it will be shown that the techniques discussed so far in \S\ref{liu}
 can be used to obtain an extended system whose only fixed eigenvalues are at 0.
 Fixed eigenvalues at 0 do not affect convergence rate for discrete-time systems. Having completed this
 discussion, attention will  be turned to the problem of achieving completely free spectrum assignment.
 This will be done in \S \ref{subsec:free-spec-state-hold} by developing a new extended system which has no fixed
 eigenvalues at all, not even eigenvalues at zero. To accomplish this, two modifications of the approach used in
\S \ref{subsec:conv-rate} are needed. The first is to transmit just the signals $x_i$ and not the signals $y_i$;
  the second is to introduce ``holding'', which comes with the price of increasing
   the dimensions of local controllers.

\vspace{-0.4cm}

\subsection{Controlled Convergence Rate} \label{subsec:conv-rate}

\vspace{-0.1cm}

Prompted by the control structure studied in the  continuous-time case, we consider a controller
   for the discrete-time multi-channel system described by \rep{mcsd} in which each agent's controller state
consists of two components, namely a sub-state $x_i\in\R^{n_i}$
 which agent $i$ communicates to its follower with possible delay, and an additional
sub-state $z_i\in R^{k_i}$ which is not shared across the network.
Agent $i$'s controller is a $(n_i+k_i)$-dimensional
linear system  with output $u_i$, state $\text{column}\{x_i,z_i\}$,
 and inputs consisting of delayed versions of the measured output $y_j$ and
  shared sub-state $x_j$
of each of its neighbors $j \in \scr{N}_i$.
We assume that each pair $(y_j,x_j)$ undergoes a delay of $d_{ij}$ integer-valued
time units when transmitted from agent $j$ to follower $i$; it is assumed that $d_{ii} =0$ for $i\in\mathbf{m}$.
Thus the signals agent $i$ receives from its neighbors  are the pairs $(y_j(t-d_{ij}), x_j(t-d_{ij}))$, $j\in\scr{N}_i$.
The state dynamics for  each $x_i$ and $z_i$  are  of the  forms
\eq{x_i(t+1) = v_i(t),\hspace{3mm} i \in \mathbf{m}\label{bunn}} and $z_i(t+1) = \nu_i(t)$
 respectively, where $v_i$ and $\nu_i$ are appropriately chosen linear functions of $z_i$
 and the signals $y_j(t-d_{ij})$, $x_j(t-d_{ij})$ for all $j\in\scr{N}_i$.

To model the delayed signals within the framework of finite-dimensional linear systems,
it is necessary to use lifting. Towards this end, for each $j\in\mathbf{m}$, let $\scr{F}_j$
 denote the set of labels of agent $j$'s followers and let
$$d_j = \max_{i\in\scr{F}_j }d_{ij}$$
For any $j\in\mathbf{m}$ for which $d_j>0$, the $d_j(n_j + q_j)$-dimensional
linear system\footnote{Here, the state $w_{j}(t) = \text{column}\{x_j(t-1)$, $x_j(t-2)$,
$\dots$, $x_j(t-d_j)$, $y_j(t-1)$, $y_j(t-2)$, $\dots$, $y_j(t-d_j)\}$.} is called a {\em lift}, which is described by the equations
\eq{w_{j}(t+1)= L_{j} w_{j}(t)+ H_{j} \begin{bmatrix} x_j(t) \cr y_j(t) \end{bmatrix} \label{lift}}
where \begin{align*}
L_j &= \text{diagonal} \left\lbrace
S_{d_{j}}\otimes I_{n_j}, S_{d_{j}}\otimes I_{q_j}
\right\rbrace \\
H_j &= \text{diagonal} \left\lbrace
e_1^{d_j}\otimes I_{n_j}, e_1^{d_j}\otimes I_{q_j}
\right\rbrace
\end{align*}
and for any positive integer $d$, $S$ is the $d$-unit shift matrix
$$S_d = \begin{bmatrix} \mathbf{0} & 0 \cr I_{d-1} & \mathbf{0} \end{bmatrix}_{d\times d}$$
Here $e_i^d$ is the $i$th unit vector in $\R^d$.
It is easy to see that if $w_j$ is properly initialized
$$\begin{bmatrix} x_j(t-d_{ij}) \cr y_j(t-d_{ij}) \end{bmatrix} = C_{ij} w_{j}(t) + D_{ij} \begin{bmatrix} x_j(t) \cr y_j(t) \end{bmatrix},\;\;i\in\scr{F}_j $$
where
\begin{align*}
C_{ij} &=
 \text{diagonal} \left\lbrace
 (e_{d_{ij}}^{d_j})' \otimes I_{n_j}, ((e_{d_{ij}}^{d_j})' \otimes I_{q_j}
 \right\rbrace \\
 D_{ij} &= 0
 \end{align*}
if $d_{ij}> 0$,
and $C_{ij} = 0$ and
$D_{ij}= I_{n_j+q_j}$
if $d_{ij} =0$.
Note that the lift described by \rep{lift} is introduced for modeling purposes only;
it is {\em not} implemented as a component sub-system of agent $j$'s controller.

In contrast to the definition of an extended system used in \S\ref{liu}, in the present setting,
 the extended system must model not only the original process \rep{mcsd} and the shared state dynamics \rep{bunn},
 but also all of  the lifts described by \rep{lift} for all $j$ for which $d_j>0$.
Towards this end, let $\tilde{x}$ denote the {\em extended state}
$$\tilde{x}  =  \text{column}\{x, x_1, w_1, x_2, w_2, \dots, x_m, w_m\}$$ and for $i\in\mathbf{m}$, define
\begin{align*}
\tilde{u}_i(t) = \text{column}\{& u_i(t), v_i(t)\}\\
\tilde{y}_i(t) = \text{column}\{& y_{j_1}(t-d_{ij_1}), y_{j_2}(t-d_{ij_2}),\ldots, \\
& y_{j_{m_i}}(t-d_{ij_{m_i}}), x_{j_1}(t-d_{ij_1}), x_{j_2}(t-d_{ij_1}), \\
& \ldots, x_{j_{m_i}}(t-d_{ij_{m_i}})\}
\end{align*}
where $\{i_1,i_2,\ldots,i_{m_{i}}\} = \scr{N}_i$.
By the {\em extended system} determined by \rep{mcsd}, \rep{bunn}, and \rep{lift} is meant the  $\tilde{n}$-dimensional, $m$-channel
linear system
\begin{align} \label{emcsdd}
&\tilde{x}(t+1) = \tilde{A}\tilde{x}(t) +\sum_{i=1}^m \tilde{B}_i\tilde{u}_i(t) \nonumber \\
&\tilde{y}_i(t) = \tilde{C}_i\tilde{x}(t), \hspace{3mm} i \in \mathbf{m}, \hspace{3mm} t \in \{0, 1, 2, \dots\}
\end{align}
where $\tilde{n} = n + \sum_{i=1}^m (n_i + d_i n_i + d_i q_i)$,
\begin{align} \label{emcsdd-mtrx}
&\tilde{A} =
\begin{bmatrix}
A & \mathbf{0} \\
G & K
\end{bmatrix}
, \nonumber \\
&\tilde{B}_i = \text{diagonal}
\{
B_i, H_{i0}
\}, \hspace{3mm} i \in \mathbf{m}, \nonumber \\
&\tilde{C}_i = \text{diagonal}
\{ C_{\scr{N}_i^0} \hspace{1mm} , \hspace{1mm}
\text{column} \{
Q_{\scr{N}_i^+} \hspace{0.5mm} , \hspace{0.5mm} H_{\scr{N}_i}'
\}
\}, \hspace{2mm} i \in \mathbf{m}
\end{align}
Here
\begin{align*}
G = \text{column}\{&G_1, G_2, \dots, G_m\} \\
G_i = \text{column}\{&\mathbf{0}_{(d_i+1)n_i \times n}, C_i, \mathbf{0}_{(d_i-1)q_i \times n}\} \\
K = \text{diagonal}\{&S_{d_1+1} \otimes I_{n_1}, S_{d_1} \otimes I_{q_1}, S_{d_2+1} \otimes I_{n_2}, \\
&S_{d_2} \otimes I_{q_2}, \dots, S_{d_m+1} \otimes I_{n_m}, S_{d_m} \otimes I_{q_m}\} \\
H_{i\delta} = \text{column} \{& \mathbf{0}_{r_i^- \times n_i}, I_{n_i}, \mathbf{0}_{r_i^+ \times n_i} \}, \hspace{3mm} 0 \leq \delta \leq d_i
\end{align*}
where $r_i^- = \delta n_i + \sum_{j=1}^{i-1}(n_j + d_j n_j + d_j q_j)$,
$r_i^+ = (d_i-\delta) n_i+d_i q_i+\sum_{j=i+1}^m (n_j + d_j n_j + d_j q_j)$; for $\mathcal{N}_i = \{j_1, j_2, \dots, j_{m_i}\}$,
$$
H_{\scr{N}_i} = \Big[\begin{matrix} H_{j_1 d_{ij_1}} & H_{j_2 d_{ij_2}} & \cdots & H_{j_{m_i} d_{ij_{m_i}}} \end{matrix}\Big] 
$$
Lastly, $\scr{N}_i^0, \scr{N}_i^+ \subset \scr{N}_i$ so that neighbor $j \in \scr{N}_i^0$ if $d_{ij} = 0$ and $j \in \scr{N}_i^+$ if $d_{ij} > 0$. Clearly, $\scr{N}_i^0 \cup \scr{N}_i^+ = \scr{N}_i$. Let
$$
Q_{i\delta} = \Big[\begin{matrix} \mathbf{0}_{q_i \times \alpha_i^-} & I_{q_i} & \mathbf{0}_{q_i \times \alpha_i^+} \end{matrix}\Big], \hspace{3mm} 1 \leq \delta \leq d_i
$$
where $\alpha_i^- = (d_i+1)n_i + (\delta-1) q_i + \sum_{j=1}^{i-1}(n_j + d_j n_j + d_j q_j)$,
$\alpha_i^+ = (d_i-\delta) q_i+\sum_{j=i+1}^m (n_j + d_j n_j + d_j q_j)$. Then for $\scr{N}_i^+ = \{j_1, j_2, \dots, j_{\gamma_i}\}$,
$$
Q_{\scr{N}_i^+} = \text{column} \left\lbrace
Q_{j_1 d_{ij_1}}, Q_{j_2 d_{ij_2}}, \cdots , Q_{j_{\gamma_i} d_{ij_{\gamma_i}}}
\right\rbrace
$$

Similar to Proposition \ref{prp:transfer-matrix-graph}, it is not hard to show that when $n_i > 0$
for all $i \in \mathbf{m}$, the union of the neighbor and transfer graphs of the
original system \rep{mcsd} is a subgraph of the transfer graph
of the above extended system \rep{emcsdd}. 
The following theorem asserts that in the face of transmission delays, the extended
  system does not have any nonzero fixed eigenvalues.
By Theorem 4 of \cite{corfmat1976decentralized},
apart from the fixed eigenvalues at 0, the rest of the closed-loop spectrum of the extended system
 can be freely assigned through any channel under the hypotheses of Theorem \ref{thm:strg-graph-delay}.
Since the convergence rate for a discrete-time system
   is determined only by the system's spectral radius, the implication of the theorem \{see Corollary \ref{cor:strg-graph-delay}\} is that
    the closed-loop extended system \{and thus the original system with delays\} can
 be exponentially stabilized at any prescribed convergence rate using distributed controllers,
  even though the system may have fixed eigenvalues at 0.

\begin{theorem} \label{thm:strg-graph-delay}
Suppose that $\mathbb{N}$ is strongly connected, that the transmission delays across $\mathbb{N}$
 are finite, and that the lower bound \rep{eqn:n_i} on $n_i$ holds for $i \in \mathbf{m}$.
 Then the extended system defined by \rep{emcsdd} has no fixed, nonzero eigenvalues.
\end{theorem}

The discussion above Theorem \ref{thm:strg-graph-delay} is summarized into the corollary below.

\begin{corollary} \label{cor:strg-graph-delay}
Under the hypotheses of Theorem \ref{thm:strg-graph-delay}, the closed-loop original system \rep{mcsd} can be exponentially stabilized at any prescribed convergence rate with distributed control.
\end{corollary}

\noindent \textbf{Proof of Theorem \ref{thm:strg-graph-delay}:} For simplicity, let $e_i$ denote the $i$th unit vector with 1 in its $i$th entry and 0 in all other entries, in which case the dimension of the space will become clear from the context. Let
\begin{equation*}
M_k \triangleq
\begin{bmatrix}
-1 & \lambda & 0 & \dots & 0 \\
0 & -1 & \lambda & \dots & 0 \\
\vdots &  & \ddots & \ddots & \vdots \\
0 & \dots & 0 & -1 & \lambda
\end{bmatrix}_{k \times (k+1)}
\end{equation*}
where $\lambda \neq 0$. It is claimed that
\begin{equation} \label{eqn:blk-rank}
\rank
\begin{bmatrix}
M_k \\
e_i'
\end{bmatrix}_{(k+1) \times (k+1)}
= k+1
, \hspace{3mm}
i \in \{1, 2, \dots, k+1\}
\end{equation}
So with proper elementary row and column operations, matrix
\begin{equation*}
\begin{bmatrix}
-C_{q \times n} & \lambda e_1' \otimes I_q \\
\mathbf{0} & M_k \otimes I_q \\
\mathbf{0} & e_i' \otimes I_q
\end{bmatrix}
, \hspace{3mm}
i \in \{1, 2, \dots, k+1\}
\end{equation*}
can be transformed into matrix $\begin{bmatrix} 
\mathbf{0} & \lambda I_{(k+1)q} \\
C & \mathbf{0} \end{bmatrix}$.

It is easy to check that the extended system \rep{emcsdd} is jointly observable but may not be jointly controllable and that the uncontrollable spectrum of $(\tilde{A}, \big[\begin{matrix} \tilde{B}_1 & \tilde{B}_2 & \dots & \tilde{B}_m \end{matrix}\big])$ consists of zeros only.

For each nonempty proper subset $\mathbf{s} \subset \mathbf{m}$, there exists $j \in \mathcal{N}_{\beta} \cap \mathbf{m} - \mathbf{s}$ for some $\beta \in \mathbf{s}$. Then by (\ref{emcsdd-mtrx}), (\ref{eqn:blk-rank}), and (\ref{eqn:n_i}), with proper elementary row and column operations, it is not hard to verify that for any $\lambda \neq 0$,
\begin{align*}
& \rank
\begin{bmatrix}
\lambda I - \tilde{A} & \tilde{B}_{\mathbf{m} - \mathbf{s}} \\
\tilde{C}_{\mathbf{s}} & \mathbf{0}
\end{bmatrix} \\
\geq \hspace{2mm} &
\rank
\begin{bmatrix}
\lambda I - A & B_{\mathbf{m} - \mathbf{s}} \\
C_{\scr{N}_{\mathbf{s}}} & \mathbf{0}
\end{bmatrix}
+
\rank
\begin{bmatrix}
\mathbf{0}        & 1 \\
M_{d_j}           & \mathbf{0} \\
e_{1+d_{\beta j}}' & 0
\end{bmatrix}
\otimes I_{n_j}
+ \\
& d_j q_j + \sum_{i \in \mathbf{m}, i \neq j} (n_i + d_i n_i + d_i q_i) \\
= \hspace{2mm} &
\rank
\begin{bmatrix}
\lambda I - A & B_{\mathbf{m} - \mathbf{s}} \\
C_{\scr{N}_{\mathbf{s}}} & \mathbf{0}
\end{bmatrix}
+
(d_j + 2)n_j + d_j q_j + \\
& \sum_{i \in \mathbf{m}, i \neq j} (n_i + d_i n_i + d_i q_i) \\
\geq \hspace{2mm} &
n + \sum_{i=1}^m (n_i + d_i n_i + d_i q_i) = \tilde{n}
\end{align*}

Thus, by Proposition \ref{prp:mtrx-pencil}, $\lambda \neq 0$ is not a fixed eigenvalue of the extended system.

Now the claim will be proved by induction on $k$. For $k=1$, $M_1 = [-1 \enspace \lambda]$. As
$
\rank
\begin{bmatrix}
-1 & \lambda \\
1  & 0
\end{bmatrix}
=
\rank
\begin{bmatrix}
-1 & \lambda \\
0  & 1
\end{bmatrix}
=2
$, the claim is true for $k=1$. Suppose the claim holds for $k=j \geq 1$, then for $i \in \{1, 2, \dots, j+1\}$,
\begin{equation*}
\rank
\begin{bmatrix}
M_{j+1} \\
e_i'
\end{bmatrix}
=
\rank
\begin{bmatrix}
M_j \\
e_i'
\end{bmatrix}
+ 1
=
j+2
\end{equation*}
Apparently,
$
\rank
\begin{bmatrix}
M_{j+1} \\
e_{j+2}'
\end{bmatrix}
=
j+2
$. So the claim also holds for $k=j+1$. This completes the proof of the claim. \hfill \qed

Theorem \ref{thm:strg-graph-delay} suggests that
the introduction of  transmission delays does not affect the bound on $n_i$ as
 long as  fixed eigenvalues at 0 are acceptable.
Similarly to Remark \ref{rmk:no-yi}, it is straightforward to verify that when only the
 signals $x_i$ and not the signals $y_i$ are transmitted across the network,
  Theorem \ref{thm:strg-graph-delay} still holds,  except in this case that the bound on the
  $n_i$ is \rep{neqn:n_i} rather than \rep{eqn:n_i}.

\vspace{-0.3cm}

\subsection{Spectrum Assignment} \label{subsec:free-spec-state-hold}

Next it will be demonstrated that with two modifications of 
the approach used in \S \ref{subsec:conv-rate}, it is possible to avoid all
 the fixed eigenvalues including those at 0 and to assign the closed-loop spectrum freely in the presence of transmission delays.

It can be checked easily that when each agent $i \in \mathbf{m}$ sends both signals $x_i$ and $y_i$ to its followers, the extended system \rep{emcsdd} may not be jointly controllable and thus may have fixed eigenvalues at 0 in the face of transmission delays. So the first modification is that each agent $i$ sends only the signal $x_i$ and not the signal $y_i$ to its followers.

The second modification is that each agent $i \in \mathbf{m}$ ``holds'' the signal $x_i$ by appropriate amount of time before sending it to each of its followers. To this end, each agent $i$ needs to know the transmission delay $d_{ji}$ from itself to each follower $j \in \mathcal{F}_i$. So each agent $i$ knows $d_i$ as well.

Roughly speaking, the idea of \emph{holding} is that instead of using the current state of itself, each agent $i$ tries to ``hold'' its state $x_i$ by appropriate amounts of time before deploying $x_i$ into its own measured output and sending $x_i$ to its other followers, so that a part of agent $i$'s own measured signal and the signals received from agent $i$ by its other followers are precisely $x_i(t - d_i)$, the maximally delayed state of agent $i$. In other words, each agent $i$ does the following. It holds its state $x_i(t)$ by $d_i$ units of time and then releases $x_i(t - d_i)$ as part of its measured output. For each of its other followers $j \in \mathcal{F}_i$, agent $i$ holds its own state $x_i(t)$ by exactly $(d_i - d_{ji})$ units of time and then transmits the state $x_i(t-d_i+d_{ji})$ to follower $j$. Taking into account the $d_{ji}$ units of delay in the transmission process, follower $j$ receives $x_i(t - d_i)$ and then makes this signal available for its own measurement. Thus all of agent $i$'s followers have $x_i(t - d_i)$ in their measured signals.

The new extended system is defined as follows. As before, each agent $i \in \mathbf{m}$ has an $n_i$-dimensional local open-loop controller
$$
x_i(t+1) = v_i(t)
$$
Let $\hat{x}$ denote the new extended state
\begin{align*}
\hat{x}(t) = \text{column} \{& x(t), x_1(t), x_1(t-1), \dots, x_1(t-d_1), \\
& x_2(t), x_2(t-1), \dots, x_2(t-d_2), \dots, \\
& x_m(t), x_m(t-1), \dots, x_m(t-d_m) \}
\end{align*}
and for $i \in \mathbf{m}$, define
\begin{align*}
\hat{u}_i(t) = \text{column} \{& u_i(t), v_i(t)\} \\
\hat{y}_i(t) = \text{column} \{& y_i(t), x_{j_1}(t-d_{j_1}), x_{j_2}(t-d_{j_2}), \dots, \\
& x_{j_{m_i}}(t-d_{j_{m_i}})\}
\end{align*}
where $\{j_1, j_2, \dots, j_{m_i}\} = \mathcal{N}_i$. Using lifting, the relationship between $\hat{u}_i(t)$ and $\hat{y}_i(t)$ can be described by a delay-free $m$-channel, $\hat{n}$-dimensional linear system of the form
\begin{align} \label{emcsdd-hold}
&\hat{x}(t+1) = \hat{A} \hat{x}(t) + \sum_{i=1}^m \hat{B}_i \hat{u}_i(t) \nonumber \\
&\hat{y}_i(t) = \hat{C}_i \hat{x}(t), \hspace{3mm} i \in \mathbf{m}, \hspace{3mm} t \in \{0, 1, 2, \dots\}
\end{align}
where $\hat{n} \triangleq n + \sum_{i=1}^m (n_i + d_i n_i)$,
\begin{align} \label{emcsdd-hold-mtrx}
\hat{A} &= \text{diagonal} \left\lbrace
A, S_{d_1+1} \otimes I_{n_1}, \dots, S_{d_m+1} \otimes I_{n_m}
\right\rbrace \nonumber \\
\hat{B}_i &= \text{diagonal} \left\lbrace
B_i, L_{i0}
\right\rbrace, \hspace{3mm} i \in \mathbf{m} \nonumber \\
\hat{C}_i &= \text{diagonal} \left\lbrace
C_i,
\left[
L_{j_1 d_{j_1}} \enspace \dots \enspace L_{j_{m_i} d_{j_{m_i}}}
\right]'
\right\rbrace
, \hspace{0.5mm} i \in \mathbf{m}
\end{align}
Here
$$
L_{i\delta} = \text{column} \left\lbrace
\mathbf{0}_{\gamma_i^- \times n_i}, I_{n_i}, \mathbf{0}_{\gamma_i^+ \times n_i} \right\rbrace, \hspace{3mm} 0 \leq \delta \leq d_i
$$
where $\gamma_i^- = \delta n_i + \sum_{j=1}^{i-1}(n_j + d_j n_j)$ and $\gamma_i^+ = (d_i-\delta) n_i+\sum_{j=i+1}^m (n_j + d_j n_j)$.

\begin{theorem} \label{thm:strg-graph-delay-hold}
Suppose that $\mathbb{N}$ is strongly connected, that the transmission delays across $\mathbb{N}$ are finite, and that the modified lower bound \rep{neqn:n_i} on $n_i$ holds for $i \in \mathbf{m}$, then the new extended system defined by \rep{emcsdd-hold} has no fixed spectrum.
\end{theorem}

\noindent \textbf{Proof of Theorem \ref{thm:strg-graph-delay-hold}:} It is easy to check that the new extended system \rep{emcsdd-hold} is jointly controllable and jointly observable.

By (\ref{emcsdd-hold-mtrx}) and \rep{neqn:n_i}, for each nonempty proper subset $\mathbf{s} \subset \mathbf{m}$ and for any $\lambda \in {\rm l \hspace{-0.45em} C}$,
\begin{align*}
\rank
\begin{bmatrix}
\lambda I - \hat{A} & \hat{B}_{\mathbf{m} - \mathbf{s}} \\
\hat{C}_{\mathbf{s}} & \mathbf{0}
\end{bmatrix}
&= \hspace{0.5mm}
\rank
\begin{bmatrix}
\lambda I_n - A & B_{\mathbf{m} - \mathbf{s}} \\
C_{\mathbf{s}} & \mathbf{0}
\end{bmatrix}
+ \\
& \hspace{5mm} \sum_{i=1}^m (n_i + d_i n_i) + \sum_{i \in \mathcal{N}_{\mathbf{s}} \cap (\mathbf{m} - \mathbf{s})} n_i \\
& \geq
\rank
\begin{bmatrix}
\lambda I_n - A & B_{\mathbf{m} - \mathbf{s}} \\
C_{\mathbf{s}} & \mathbf{0}
\end{bmatrix}
+ \\
& \hspace{5mm} \sum_{i=1}^m (n_i + d_i n_i) + \min_{i \in \mathbf{m}} \, n_i \\
& \geq
n + \sum_{i=1}^m (n_i + d_i n_i) = \hat{n}
\end{align*}

Therefore, by Proposition \ref{prp:mtrx-pencil}, the new extended system has no fixed spectrum. \hfill \qed

Similar to Proposition \ref{prp:transfer-matrix-graph}, it is straightforward to verify that when $n_i > 0$ for all $i \in \mathbf{m}$, the transfer graph of the new extended system (\ref{emcsdd-hold}) is strongly connected if and only if the union of the neighbor and transfer graphs of the original system (\ref{mcsd}) is strongly connected. Therefore, under the hypotheses of Theorem \ref{thm:strg-graph-delay-hold}, the closed-loop spectrum of the new extended system can be freely assigned via a suitably designed channel controller $z_q$ \cite{brasch1970pole} for any selected channel $q \in \mathbf{m}$ \cite{corfmat1976decentralized}. Other yet to be developed design techniques may possibly also be applied.

\vspace{-0.3cm}

\section{Concluding Remarks}

This paper presents new systematic procedures for
 exponentially stabilizing a jointly controllable, jointly observable, multi-channel linear system
  with any prescribed convergence rate using linear time-invariant distributed controllers.
  While the first procedure in \S\ref{sec:dis-obs-contr} addresses the problem using a distributed observer,
  the second in \S \ref{liu} casts the problem in a more general setting which is broad enough
to enable alternative and yet to be devised
 solutions. There is a large literature focusing on
 decentralized control from an optimization perspective
 \{e.g., \cite{rotkowitz2006characterization}\}, which may prove useful here.

This paper also draws attention to the practical but challenging  problem
of dealing with transmission delays across a network. It would be interesting to see  how ideas from the
 literature on control of
linear systems over networks  with limited data rates \{e.g., \cite{nair2003exponential, tatikonda2004control}\}
might be brought to bear on this problem. Other issues such as the effects of asynchronous operations
 and resilience to local failures would also be of interest to study.

 Practical feedback control procedures have features not really possessed by the algorithms proposed in this   paper
or by the algorithms described in \cite{kexin,plug,Luis,XZ}.
 None of the
algorithms address design metrics \{other than convergence rate\}, potential problems with unmodeled dynamics,  noise,
the range over which a linear process model  is valid, etc. In our view all of these algorithms should be thought
of as preliminary ideas contributing to what might at some point become
 a bona fide  practical feedback theory  for distributed systems.
 It was with this type of thinking in
 mind that we were inspired to
develop the material in \S \ref{liu} to demonstrate how any solution to the basic  distributed control
problem could be viewed through the lens of decentralized control technology. There is considerable value in doing this.
 For example, Theorem \ref{thm:strg-graph}
 gives a clear indication
 of the  dimensions  any  distributed controllers must    satisfy  in order to provide solutions to the
 basic problem. Moreover, this setup
 enables one to draw some conclusions
 about what type of shared signals will and will not enable solutions to the basic problem.
  An instance of this is the example in \S \ref{nope} which shows quite surprisingly,
  that if the only signal each agent transmits to its neighbors
  is its measured output $y_i$, then
  distributed stabilization may be impossible, no matter how much local dynamics each agent uses.
  It is  results like this which are needed to enable a distributed feedback control theory.


\bibliographystyle{IEEEtran}
\bibliography{Group-TAC}

\end{document}